\documentclass[preprint,pdf]{iucr} 
  \journalcode{J}
  \papertype{FA} 
  
                \def\href#1{\relax}\let\foo\caption
\ifPDF
\RequirePackage{hyperref}
\PassOptionsToPackage{pdftex,bookmarksopen,bookmarksnumbered}{hyperref}
\fi
\let\caption\foo

\usepackage{amsmath}
\usepackage{graphicx}
\graphicspath{{figures/}}
\usepackage{tikz}
\usepackage{standalone}
\usepackage{siunitx}

\usepackage{threeparttable}
\usepackage{algpseudocode}

\newcommand\inputpgf[2]{{
		\let\pgfimageWithoutPath\pgfimage
		\renewcommand{\pgfimage}[2][]{\pgfimageWithoutPath[##1]{#1/##2}}
		\let\includegraphicsWithoutPath\includegraphics
		\renewcommand{\includegraphics}%
		[2][]{\includegraphicsWithoutPath[##1]{#1/##2}}
		\input{#1/#2}
}}

\begin{document}                  



\title{Automatic indexing of wide-band quasi-Laue neutron diffraction patterns
of large unit cells}
\shorttitle{Indexing wide-band quasi-Laue neutron patterns}


\cauthor[a,b]{Zachary}{Morgan}{zmorgan@kettering.edu}{}
\author[c]{Nathan}{Whittington}
\author[c]{Iris}{Ye}
\author[d]{Fiona}{He}
\author[e]{Jacob}{Hauck}
\author[f]{Khadiza}{Begam}
\author[g]{Guannan}{Zhang}
\author[f]{Jens}{Glaser}
\author[a]{Christina}{Hoffmann}
\author[a]{Dean}{Myles}

\aff[a]{Neutron Scattering Division, Oak Ridge National Laboratory}
\aff[b]{Department of Mechanical Engineering, Kettering University}
\aff[c]{Next Generation Pathways to Computing at ORNL}
\aff[d]{Next Generation STEM Internship Program Participant at ORNL}
\aff[e]{Department of Mathematics and Statistics, Missouri University of
Science and Technology}
\aff[f]{National Center for Computational Sciences, Oak Ridge National
Laboratory}
\aff[g]{Computer Science and Mathematics Division, Oak Ridge National
Laboratory}


\shortauthor{Morgan {\it et al.}}







\maketitle                        

\begin{synopsis}
An algorithm for automatic indexing of quasi-Laue neutron diffraction patterns
with a known unit cell and a wide wavelength band combines a global orientation
search, an exact $k$-d tree search of candidate reflections and a chance-match
score. It is validated against time-of-flight indexing and applied
to the new IMAGINE-X instrument.
\end{synopsis}

\begin{abstract}
With a continuous spectrum of incident neutron wavelengths, quasi-Laue
diffraction records many Bragg reflections from a stationary crystal, each
satisfying the diffraction condition at a particular wavelength. Because the
wavelength of each reflection is not measured, the crystal orientation must be
determined from the scattering directions alone. At the
new IMAGINE-X instrument, instant-readout detectors and many short exposures
per experiment require this indexing to run automatically, yet for a protein
crystal at the full 2--10~\AA{} band a single peak can have $\sim10^4$
candidate reflections, and a random orientation indexes a large fraction of
the observed peaks. We describe an algorithm that indexes such patterns for a
known unit cell and wavelength band. Differential evolution searches uniformly
sampled rotations, matching observed directions to band-limited shells of
candidate directions, and each solution is tested against the indexing
expected for a random orientation. In a benchmark search, a $k$-d tree and a
persistent pool of 32 worker processes reduce the wall time 270-fold relative
to a serial scan of all candidates. The
algorithm is first validated on the time-of-flight diffractometer MANDI, where
the measured wavelengths give a reference solution by standard indexing. With
the wavelengths discarded, every indexed reflection from 93 rare-earth garnet
patterns agrees up to harmonic order. The algorithm then reliably indexes the
natural mineral mesolite, measured on the original image plate at IMAGINE
before the upgrade, and the protein T4 lysozyme, measured on the new IMAGINE-X
instrument over the full band.
\end{abstract} \bigskip\bigskip


\section{Introduction} \bigskip

The first observation of X-ray diffraction by a crystal, recorded in 1912 by
Friedrich, Knipping and Laue \cite{friedrich_interferenzerscheinungen_1913},
was made with the unfiltered radiation of an X-ray tube.
Its spots were initially attributed to a few discrete wavelengths present in
the incident beam. W.~L.~Bragg showed instead that the beam has a continuous
spectrum, from which each set of lattice planes of spacing $d$ reflects only
the wavelengths $\lambda$ that satisfy $n\lambda=2d\sin\theta$ at the
scattering angle $2\theta$ \cite{bragg_diffraction_1913,bragg_reflection_1913}.
A stationary crystal in a polychromatic beam therefore records many Bragg
reflections at once, but not the wavelength, or the harmonic order $n$, of any
of them. Identifying the reflection behind each spot without its wavelength has
remained the central difficulty of indexing Laue patterns ever since.

With modern position-sensitive area detectors
\cite{helliwell_recording_1989,mcintyre_area_2015}, Laue diffraction remains a
powerful technique for rapid single-crystal measurements across X-ray, neutron,
and electron diffraction. Its ability to collect many reflections in a single
exposure makes efficient use of the available flux and measurement time, which
is particularly useful for weakly scattering samples, small crystals and
neutron experiments \cite{myles_neutron_1997,meilleur_imagine_2020}. Recording
a full pattern in a single exposure also enables time-resolved studies
\cite{moffat_time-resolved_2001}.
Neutron diffraction provides sensitivity to hydrogen and deuterium positions,
making it possible to identify protonation states, hydrogen-bonding networks,
and chemically active sites in biological structures
\cite{niimura_neutron_2011,meilleur_neutron_2018}. Dynamic nuclear polarization
can further enhance the contrast from hydrogen
\cite{pierce_chapter_2020,begam_hydrogen_2025}.

At Oak Ridge National Laboratory, single-crystal neutron diffraction
experiments are performed using both continuous and pulsed neutron sources
\cite{coates_suite-level_2018}. The High Flux Isotope Reactor (HFIR) provides a steady
neutron flux, whereas the Spallation Neutron Source (SNS) provides a time-structured,
pulsed neutron beam suitable for time-of-flight measurements, in which the
flight time of each detected neutron gives its wavelength. As a result, observed detector-space
peak positions can be converted directly into reciprocal-space scattering
vectors, enabling automated determination of the UB matrix, which describes the
lattice and its orientation relative to the instrument, in software packages such
as Mantid \cite{arnold_mantiddata_2014} and NeuXtalViz
\cite{morgan_neuxtalviz_2026}, building on earlier tools developed for
pulsed-source neutron diffraction \cite{jacobson_orientation-matrix_1986}.

Laue diffraction at continuous-source instruments presents a different indexing
problem. At IMAGINE, a macromolecular crystallography beamline at HFIR
\cite{meilleur_imagine_2013,meilleur_imagine_2020}, the wavelength associated
with each observed Laue spot cannot be obtained directly. The detector position therefore
constrains the scattered-beam direction, but not the full, unique reciprocal-space
vector. Indexing therefore requires finding the crystal orientation for which
reflections within the accessible wavelength band are predicted at the
observed spots. For the image-plate detector
of IMAGINE, indexing and integration are done with LAUEGEN, adapted for
neutron diffraction and cylindrical detectors
\cite{campbell_lauegen_1995,meilleur_imagine_2020}. LAUEGEN finds the
orientation for a known cell by matching the angles between observed and
predicted low-index spots, and the user selects and refines a solution with an
interactive display of the predicted and observed patterns
\cite{campbell_lauegen_1995}.

IMAGINE is currently being upgraded to IMAGINE-X, which replaces the
cylindrical image plate with up to 80 instant-readout Anger cameras
\cite{loyd_sub-millimeter_2024} in two arrays of 40. Furthermore, it adds
dynamic nuclear polarization of the sample
\cite{pierce_chapter_2020,begam_hydrogen_2025}, with a cryogenic neutron spin
flipper \cite{fu_high-efficiency_2026}. Before and after the upgrade, the
instrument uses the same beamline optics on cold guide CG-4D of HFIR, which
delivers neutrons over 2--10~\AA. Exchangeable flat mirrors and filters set a
short-wavelength cutoff of 2.0, 2.8 or 3.3~\AA{} and a long-wavelength cutoff
of 3.0, 4.0 or 4.5~\AA, and a pair of elliptical mirrors focuses the beam to
$2\times3.2$~mm$^2$ at the sample
\cite{meilleur_neutron_2018,meilleur_imagine_2020}. The standard quasi-Laue
window is 2.8--4.5~\AA. With the long-wavelength cutoff left open, as for the
T4 lysozyme data here, the band extends to the full 2--10~\AA{} spectrum of the
guide, and a short exposure records many reflections. This open mode is used
to assess the crystal quality and confirm the unit cell efficiently before data
are collected in the quasi-Laue window, and the orientation from such an image
can be used to plan the crystal settings for the rest of the experiment
\cite{meilleur_imagine_2020,morgan_neuxtalviz_2026}. This requires indexing a
single exposure on its own, without a rotation series. Polarization
experiments need short exposures from many crystal orientations, and the
instant-readout detectors make each exposure available as it is collected.
Indexing therefore has to run without user intervention, be reliable for
protein unit cells over the full band, and keep up with data collection.

Indexing of diffraction patterns has a large literature
\cite{morawiec_indexing_2022}, and many of its ideas carry over to this
problem. Autoindexing of monochromatic oscillation data
\cite{steller_algorithm_1997,powell_autoindexing_2013,winter_dials_2018} is
well established and uses the known wavelength to obtain scattering vectors
directly. For single-crystal Laue data without wavelength information, an
early semi-automatic method recognized zone conics to construct candidate
cells \cite{ravelli_towards_1996}, and ab initio methods determine the cell
without prior knowledge \cite{morawiec_ab_2021}. LAUEGEN, developed for
synchrotron X-ray Laue data \cite{campbell_lauegen_1995}, has been used to
index neutron quasi-Laue data from macromolecules on cylindrical image plates
since LADI \cite{myles_neutron_1997,meilleur_imagine_2020}. On neutron Laue
diffractometers at reactor sources
\cite{ouladdiaf_orientexpress_2006,ouladdiaf_cyclops_2011}, LaueG obtains an
initial orientation from a user-supplied cell by automated spot detection and
indexing, with graphical tools to inspect the solution, change the lattice or
guide the indexing manually \cite{piltz_laueg_2018}, before the orientation is
refined and the intensities corrected \cite{piltz_accurate_2018}. LauePt4 finds
the orientation of a single crystal from the interplanar angles of spots
selected interactively \cite{huang_upgraded_2023}. Laue-DIALS indexes
polychromatic X-ray data from macromolecular crystals in two steps
\cite{hewitt_laue-dials_2024}. Monochromatic autoindexing at a nominal
wavelength, applied to the reflections of all images in a series, gives the
cell and an initial orientation. Miller indices and wavelengths within the band
are then assigned by matching observed and predicted scattering directions,
while the orientation of each image is refined.

Related methods address grain mapping and serial crystallography.
In X-ray microdiffraction, angular matching of reflection pairs with a known
unit cell identifies grain orientations
\cite{chung_automated_1999,tamura_scanning_2003}.
A dictionary--branch--bound search indexes superimposed Laue patterns from
many grains \cite{seret_indexing_2022}.
LaueMatching instead correlates images directly with a precomputed library
on a dense orientation grid, avoiding peak finding
\cite{sharma_lauematching_2026}.
Neural networks trained on simulated patterns recognize Laue spot indices
\cite{purushottam_raj_purohit_lauenn_2022}, while physics-informed machine
learning enables mapping of grains smaller than the beam
\cite{chan_physics-informed_2025}.
For neutrons, forward-model approaches recover grain positions and
orientations from Laue patterns recorded at multiple sample orientations
\cite{raventos_laue_2019,kacprzak_laue_2024}.

Serial crystallography requires independent indexing of diffraction snapshots
from many crystals.
Pair-angle matching combined with nearest-neighbor matching indexes both
Laue and monochromatic serial data \cite{dejoie_pattern-matching_2020}.
pinkIndexer indexes polychromatic snapshots, including pink-beam
macromolecular data \cite{gevorkov_pinkindexer_2020}.
For monochromatic serial data, XGANDALF uses extended gradient descent
\cite{gevorkov_xgandalf_2019}, TORO uses robust optimization
\cite{gasparotto_toro_2024}, and a recent method searches orientations
globally using differential evolution \cite{nasser_robust_2025}.

On IMAGINE-X, each image comes from a single crystal of known cell, as in
single-crystal Laue diffraction, but has to be indexed unattended, from a single
exposure, as it is collected, as in serial crystallography. The band can also span a factor of
five in wavelength, so a high-angle peak of a protein crystal can have
$\sim10^4$ candidate reflections. With so many candidates, a random orientation indexes a large
fraction of the observed peaks, so a solution has to be judged against the
indexing expected by chance.

Here we describe an indexing algorithm for this regime. Given a known cell and
wavelength band, it finds the crystal orientation by a global search over
orientations that compares observed and predicted scattering directions,
without assigning wavelengths in advance. The main developments are:
(i) a formulation of quasi-Laue indexing as matching of directions against
band-limited candidate shells, with harmonics collapsed to one direction per
reciprocal-lattice row;
(ii) a chance-match score that compares a solution with the indexing expected
for a random orientation, which also selects the peaks that carry information about
the orientation;
(iii) an exact nearest-candidate search with a $k$-d tree and a persistent pool
of worker processes, which together shorten a search for a protein crystal lattice by
more than two orders of magnitude; and
(iv) validation against time-of-flight indexing on MANDI and the first indexing
of data from IMAGINE-X, including the detector geometry calibration this
required.

The algorithm is evaluated using diffraction data from mesolite, rare-earth
garnet and T4 lysozyme crystals collected on the IMAGINE, MANDI and IMAGINE-X
instruments. We conclude with the limitations of the method and its place in
the data reduction for IMAGINE-X.

\section{Quasi-Laue diffraction geometry} \bigskip

This section sets out the geometry on which the indexing algorithm is built. We
first fix the laboratory and reciprocal-space conventions, then write the Laue
condition for a peak whose wavelength is unknown, and show how the wavelength
band limits the reflections that can produce each peak. Two consequences of the
missing wavelength, an undetermined cell scale and harmonic reflections that
share a direction, shape the algorithm of Section~\ref{sec:algorithm}.

\subsection{Conventions}

Laboratory coordinates follow Mantid \cite{arnold_mantiddata_2014}. The
incident beam travels along $+z$, $y$ points vertically up and horizontal $x$ completes a
right-handed frame. A detected neutron of wavelength $\lambda$, scattered by the scattering (polar) angle $2\theta$
(with $\theta$ the Bragg angle) at azimuth $\phi$, travels along the
direction $\hat{\boldsymbol{k}}_f$ of its final (scattered) wavevector, while the
incident wavevector points along $\hat{\boldsymbol{k}}_i$,
\begin{equation}
\hat{\boldsymbol{k}}_f=
\begin{bmatrix}
\sin2\theta\cos\phi\\
\sin2\theta\sin\phi\\
\cos2\theta
\end{bmatrix},
\qquad
\hat{\boldsymbol{k}}_i=
\begin{bmatrix}
0\\0\\1
\end{bmatrix},
\label{eq:kf}
\end{equation}
and detector views are displayed with the horizontal and vertical angles
$\gamma=\operatorname{atan2}(x,z)$ and $\nu=\arcsin(y/|\boldsymbol{r}|)$ of the
pixel position $\boldsymbol{r}$.
We use the crystallographic convention without the factor $2\pi$. A reflection
$\boldsymbol{h}=(h,k,l)^\mathsf{T}$ has scattering vector
\begin{equation}
\boldsymbol{s}=h\boldsymbol{a}^\ast+k\boldsymbol{b}^\ast+l\boldsymbol{c}^\ast=
\boldsymbol{R}\boldsymbol{U}\boldsymbol{B}\boldsymbol{h},\qquad
|\boldsymbol{s}|=1/d_{hkl},
\label{eq:ub}
\end{equation}
where $\boldsymbol{a}^\ast$, $\boldsymbol{b}^\ast$, $\boldsymbol{c}^\ast$ are
the reciprocal basis vectors, $d_{hkl}$ is the interplanar spacing,
$\boldsymbol{B}$ is the upper-triangular reciprocal-lattice matrix, the
Cholesky factor of the reciprocal metric tensor
$\boldsymbol{G}^\ast=\boldsymbol{B}^\mathsf{T}\boldsymbol{B}$, $\boldsymbol{U}$
is the orientation of the crystal on the goniometer and $\boldsymbol{R}$ is the
goniometer rotation \cite{busing_angle_1967}. The momentum transfer of Mantid
is $\boldsymbol{Q}=2\pi\boldsymbol{s}$ (with \texttt{Q.convention} set to
\texttt{Crystallography}).

\subsection{Laue condition and wavelength ambiguity}

\begin{figure}
	\includegraphics[width=\textwidth]{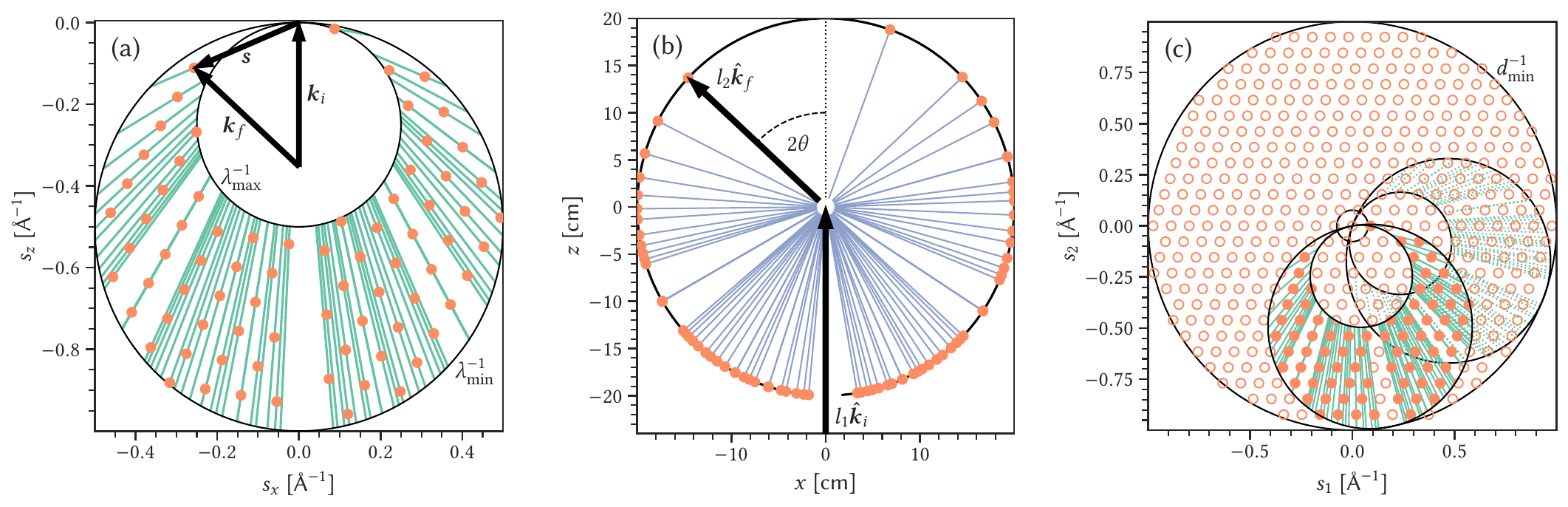}\\
	\caption{
		Geometry of quasi-Laue diffraction in the horizontal plane.
		(a) Ewald construction for the band
		$\lambda_\text{min}\le\lambda\le\lambda_\text{max}$: reciprocal-lattice
		points (orange) between the Ewald spheres of radius $\lambda_\text{max}^{-1}$
		and $\lambda_\text{min}^{-1}$ diffract, with scattering vector
		$\boldsymbol{s}=\boldsymbol{k}_f-\boldsymbol{k}_i$ for the incident and final
		wavevectors $\boldsymbol{k}_i$ and $\boldsymbol{k}_f$ of length $\lambda^{-1}$.
		Lines through the origin
		(green) are reciprocal-lattice rows, whose harmonics $n\boldsymbol{h}$
		diffract into the same direction.
		(b) The same reflections in laboratory coordinates, with scattered rays (blue)
		from the sample at scattering angle $2\theta$ to the detector (orange) at
		distance $l_2$, for the incident beam $\hat{\boldsymbol{k}}_i$.
		(c) Reciprocal lattice within $d_\text{min}^{-1}$ in the crystal frame (open
		circles) and the region covered by the band for two orientations, with the reflections in the band for the actual orientation of the crystal filled.
	}
	\label{fig:laue}
\end{figure}

For each observed peak $i$ the detector position gives
\begin{equation}
\boldsymbol{q}_i=\hat{\boldsymbol{k}}_f-\hat{\boldsymbol{k}}_i=2\sin\theta_i
\begin{bmatrix}
\cos\theta_i\cos\phi_i\\
\cos\theta_i\sin\phi_i\\
-\sin\theta_i
\end{bmatrix},
\label{eq:q}
\end{equation}
the dimensionless scattering vector of the peak, with
$|\boldsymbol{q}_i|=2\sin\theta_i$. Bragg's law in vector form,
$\boldsymbol{s}=\lambda^{-1}(\hat{\boldsymbol{k}}_f-\hat{\boldsymbol{k}}_i)$,
reads
\begin{equation}
\boldsymbol{q}_i=\lambda_i\,\boldsymbol{R}\boldsymbol{U}\boldsymbol{B}
\boldsymbol{h}_i,
\qquad \lambda_\text{min}\le\lambda_i\le\lambda_\text{max}.
\label{eq:laue}
\end{equation}
Without time-of-flight, $\lambda_i$ is unknown, so a peak fixes only the
direction of $\boldsymbol{s}$:
\begin{equation}
\hat{\boldsymbol{q}}_i=
\begin{bmatrix}
\cos\theta_i\cos\phi_i\\
\cos\theta_i\sin\phi_i\\
-\sin\theta_i
\end{bmatrix}
=d_{hkl}\,\boldsymbol{R}\boldsymbol{U}\boldsymbol{B}\boldsymbol{h}_i,
\qquad
\lambda_i=2d_{hkl}\sin\theta_i=\frac{|\boldsymbol{q}_i|}{|\boldsymbol{B}
\boldsymbol{h}_i|}.
\label{eq:direction}
\end{equation}
The band restricts the reflections that can produce peak $i$ to the reciprocal-space shell
\begin{equation}
\mathcal{S}_i=\left\{\boldsymbol{h}:\
\frac{2\sin\theta_i}{\lambda_\text{max}}\le|\boldsymbol{B}\boldsymbol{h}
|\le\frac{2\sin\theta_i}{\lambda_\text{min}}\right\}
\label{eq:shell}
\end{equation}
of $N_i$ candidates (Fig.~\ref{fig:laue}). Harmonics $n\boldsymbol{h}$ are
collinear with $\boldsymbol{h}$ and diffract into the same direction at
$\lambda/n$, so the shell contains $n_i\le N_i$ distinct directions, one per
primitive reciprocal-lattice row \cite{cruickshank_multiplicity_1987}.

Two consequences shape the algorithm. First, scaling every cell length by $t$
scales $|\boldsymbol{B}\boldsymbol{h}|$ by $1/t$, which is absorbed by
$\lambda_i\to t\lambda_i$ in Eq.~(\ref{eq:laue}), so the absolute cell scale is
not determined, only length ratios and angles. In the cubic case
$\boldsymbol{B}=\boldsymbol{I}/a$ and Eq.~(\ref{eq:direction}) becomes
$\hat{\boldsymbol{q}}_i=\boldsymbol{R}\boldsymbol{U}\hat{\boldsymbol{h}}_i$,
independent of $a$. Second, the number of candidate directions grows with
$\sin^3\theta_i$ and with the cell volume, so high-angle peaks of large cells
can be matched by almost any orientation.

\section{Indexing algorithm} \bigskip
\label{sec:algorithm}

Laue indexing has to determine the Miller indices and the wavelength of every
reflection at once, since many reciprocal-lattice vectors may satisfy a given
scattering direction \cite{chung_automated_1999}. Autoindexing methods for
monochromatic oscillation data
\cite{steller_algorithm_1997,powell_autoindexing_2013,winter_dials_2018} rely
on a known wavelength. For polychromatic data, approaches include ab initio
indexing of Laue patterns \cite{morawiec_ab_2021} and pink-beam indexing
\cite{gevorkov_pinkindexer_2020}. For large unit cells such as those of
proteins, combinatorial matching must consider many candidate reflections per
peak, missing and spurious peaks add to the combinations, and the orientation
search space has many local minima \cite{ren_laue_1999}. In quasi-Laue diffraction the
restricted band reduces, but does not remove, the coupling between wavelength
and indexing
\cite{wilkinson_quasi-laue_1991,myles_neutron_1997,meilleur_neutron_2006,%
blakeley_neutron_2010}.
We therefore treat indexing as a global optimization over orientation for a
known cell and band, using differential evolution (DE)
\cite{storn_differential_1997}, a population-based global optimizer that, like
particle swarm optimization \cite{kennedy_particle_1995}, needs neither
gradients nor an initial guess. Miller indices and wavelengths are assigned
implicitly for each trial orientation, with the band acting as a constraint
(Eq.~\ref{eq:shell}).

\subsection{Peak finding and scattering directions}

Each camera image is background-subtracted with a median filter, and its noise
$\sigma$ is estimated from the median absolute deviation of the residual \cite{weng_k-space_2020}. Row
and column medians are then subtracted to remove line artifacts. The residual
$r$ is convolved with a normalized Gaussian kernel $w$ of the spot size (one
pixel) and divided by the noise of the result, giving the significance
$z=(w\ast r)/(\sigma\lVert w\rVert_2)$, where $\ast$ denotes convolution and
$\lVert w\rVert_2$ the Euclidean norm of the kernel. Peaks are local maxima
with $z\ge5$, located by the residual-weighted centroid. A matched filter
thresholds the summed signal of a spot rather than its brightest pixel, which
matters for the faint spots that dominate at high scattering angles. Pixel
positions, including any calibration, give $\boldsymbol{q}_i$ through
Eq.~(\ref{eq:q}), and $\boldsymbol{R}^\mathsf{T}\boldsymbol{q}_i$ in the sample
frame.

\subsection{Orientation parameterization}
\label{sec:orientation}

The candidate orientation matrix $\boldsymbol{U}$ is parameterized by a rotation $\omega$ about the unit axis
$\hat{\boldsymbol{n}}=(\sin\vartheta\cos\varphi,$ $\sin\vartheta\sin\varphi,$
$\cos\vartheta)$,
\begin{equation}
\boldsymbol{U}=\boldsymbol{I}+\sin\omega\,[\hat{\boldsymbol{n}}]_\times+
(1-\cos\omega)\,[\hat{\boldsymbol{n}}]_\times^2,
\label{eq:rodrigues}
\end{equation}
with $[\hat{\boldsymbol{n}}]_\times$ the cross-product matrix (written out in
Appendix~\ref{app:rotation}). The optimizer works on $(u_0,u_1,u_2)\in[0,1]^3$
with
\begin{equation}
\cos\vartheta=1-2u_0,\qquad \varphi=2\pi u_1,\qquad
\frac{\omega-\sin\omega}{\pi}=u_2,
\label{eq:uniform}
\end{equation}
so that $0\le\vartheta\le\pi$, $0\le\varphi<2\pi$ and $0\le\omega\le\pi$. The
axis is uniform on the sphere and $\omega$ follows the density
$(1-\cos\omega)/\pi$, so a uniform draw of $(u_0,u_1,u_2)$ is a uniform random
rotation \cite{miles_random_1965} (Appendix~\ref{app:sampling}). The DE
initial population therefore covers orientation space without bias, while
later generations concentrate near good solutions.

\subsection{Objective function}

For a trial $\boldsymbol{U}$, the angular error of peak $i$ is the angle to the
nearest candidate direction in its shell,
\begin{equation}
\delta_i(\boldsymbol{U})=\min_{\boldsymbol{h}\in\mathcal{S}_i}
\arccos\left[\left(\boldsymbol{U}^\mathsf{T}\boldsymbol{R}^\mathsf{T}
\hat{\boldsymbol{q}}_i\right)\cdot\frac{\boldsymbol{B}\boldsymbol{h}}
{|\boldsymbol{B}\boldsymbol{h}|}\right],
\label{eq:delta}
\end{equation}
which assigns $\boldsymbol{h}_i$ and, through Eq.~(\ref{eq:direction}),
$\lambda_i$. Unlike a residual in fractional indices,
$\boldsymbol{h}-\operatorname{round}(\boldsymbol{h})$, Eq.~(\ref{eq:delta})
does not require $\lambda_i$, which is unknown. The objective is
\begin{equation}
C(\boldsymbol{U})=\sum_i
w_i\min\left(\frac{\delta_i}{\tau},\,z_\text{max}\right)^2-\beta\sum_{i:\,
\delta_i\le\tau}w_i,
\label{eq:cost}
\end{equation}
a capped quadratic in $\delta_i/\tau$ with a bonus $\beta$ for every peak
indexed within the tolerance $\tau$ ($z_\text{max}=3$, $\beta=4$). Matching
observed directions to lattice directions by a rotation, with spurious peaks
present, is the constellation problem, and Eq.~(\ref{eq:cost}) is a
trial-and-error objective of the kind reviewed by \citeasnoun{morawiec_indexing_2022},
which requires a term that favors solutions with many matched peaks, here the bonus $\beta$. The weights
$w_i\propto\sqrt{I_i}/\sqrt{N_i}$, normalized to unit mean, favor strong peaks
with few candidates, where $I_i$ is the background-subtracted peak height
clipped to its 10th--95th percentiles. The weights are capped at twice their
mean so that no single peak dominates the objective. Peaks close to the
direct beam have the fewest candidates and would otherwise receive the largest
weights, although spurious features from the direct beam can also be detected
there.

Evaluating Eq.~(\ref{eq:delta}) by scanning every candidate costs $\sum_iN_i$
dot products per trial orientation, which reaches $10^6$--$10^7$ for proteins
at a wide band. The cost can be reduced with a nearest-neighbor search. Each
candidate is represented by its direction
$\hat{\boldsymbol{g}}=\boldsymbol{B}\boldsymbol{h}/|\boldsymbol{B}
\boldsymbol{h}|$
in the Cartesian frame of the crystal, and each peak by its observed direction
rotated into the same frame,
$\boldsymbol{U}^\mathsf{T}\boldsymbol{R}^\mathsf{T}\hat{\boldsymbol{q}}_i$. The
distance between two unit vectors separated by an angle $\delta$ is
$2\sin(\delta/2)$, which increases monotonically with $\delta$, so the
candidate at the smallest angle is also the nearest neighbor of the rotated
peak direction. Nearest neighbors can be found quickly in a $k$-d tree
\cite{bentley_multidimensional_1975}, which is built once, before the search,
because the candidate directions do not depend on the orientation.

\begin{figure}
	\includegraphics[width=\textwidth]{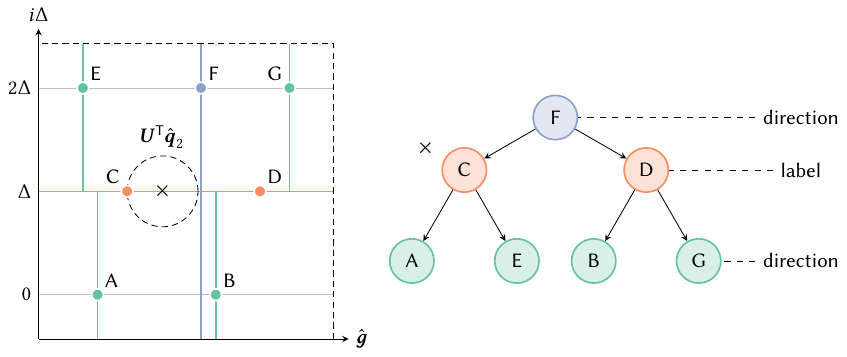}\\
	\caption{
		Labeled $k$-d tree for the nearest-candidate search (schematic, not to
		scale). Left: the candidate directions $\hat{\boldsymbol{g}}$ of each peak
		are placed in a layer at height $i\Delta$ (here three peaks), and the space
		is split alternately along the direction and label coordinates. Right: the
		corresponding tree. The query for peak $i$ is its rotated observed direction
		$\boldsymbol{U}^\mathsf{T}\hat{\boldsymbol{q}}_i$ in layer $i\Delta$
		($\times$, peak 2). Because the layer spacing $\Delta$ exceeds the largest
		chord between directions, its nearest neighbor (dashed circle) is always one
		of the candidates of the same peak (C).
	}
	\label{fig:kdtree}
\end{figure}

One restriction must be respected. Each peak may be matched only by the
candidates in its own shell $\mathcal{S}_i$, since a reflection from another
shell would require a wavelength outside the band at that scattering angle. A
single tree of all candidate directions would ignore this, and a separate tree
for each peak would require hundreds of separate searches per trial
orientation. Instead, we add a fourth coordinate that records which peak a
candidate belongs to. Every candidate of peak $i$ is stored as the point
$(\hat{\boldsymbol{g}},\,i\Delta)$ and the peak is queried at
$(\boldsymbol{U}^\mathsf{T}\boldsymbol{R}^\mathsf{T}\hat{\boldsymbol{q}}_i,\,
i\Delta)$,
where $i$ is the index of the peak in the peak list and $\Delta$ is a constant.
The fourth coordinate is a dimensionless label with no physical meaning. It places the candidates of each peak in a separate layer of the four-dimensional
space (Fig.~\ref{fig:kdtree}). Within a layer, two points are at most 2 apart,
the largest distance between unit vectors, whereas points in different layers
are at least $\Delta$ apart. With $\Delta>2$ (we use $\Delta=4$), the nearest
neighbor of the query for peak $i$ is therefore always one of its own
candidates, exactly as if peak $i$ had a tree of its own. All peaks are then
searched in a single call to one tree, which gives the same minimum as the scan
at a fraction of the cost.

\subsection{Global search}
\label{sec:search}

The cost $C$ is minimized over $(u_0,u_1,u_2)$ by DE, which has also been used to search
orientations in monochromatic serial indexing \cite{nasser_robust_2025}, with
the \texttt{rand/1/bin} strategy, 300 population members per parameter (900,
rounded up to 1024 by the Sobol initialization), dithered mutation in
$[0.5,1.2]$, crossover probability 0.5, up to 1000 generations and no final
gradient polishing \cite{storn_differential_1997,virtanen_scipy_2020}. Each
generation is evaluated in parallel by worker processes (Algorithm~1).
Only informative peaks are used. A peak whose chance of being matched by
a random orientation exceeds 0.7 at the final tolerance is likely to match
even at an incorrect orientation and therefore provides little discrimination, so it is indexed only once the orientation is known. This chance is evaluated with
Eq.~(\ref{eq:chance}), conservatively for all $N_i$ candidates rather than the
$n_i$ distinct directions, so that peaks with many harmonics are also
excluded. From the informative peaks, the search uses a subset of up to 80.
The peaks are ranked by $w_i/\sqrt{N_i}$, which favors strong peaks with few
candidates, and taken in order, skipping any within $1^\circ$ of a peak already
chosen so that the subset is spread over the detector. If fewer than 80 remain,
the subset is filled by rank. Peaks with more than $2\times10^5$ candidates are
left out. The search uses a wide tolerance, $\tau_s=1^\circ$, which broadens
the basin of the true orientation. Its result is refined by assignment and
alignment (Section~\ref{sec:refine}) in two cycles on the subset and three on
all informative peaks at $\tau_s$, followed by three cycles at the final
tolerance $\tau=0.35^\circ$. Each attempt is scored by the chance-match score
(Section~\ref{sec:significance}), taking the more significant of its values at
$\tau$ and $\tau_s$. Up to three attempts are made with new seeds, stopping at
the first that reaches $\log_{10}p<-5$. If none does, the most significant
attempt is returned with its score, and the pattern is reported as not
indexed.
Differential evolution was chosen over particle swarm optimization
\cite{kennedy_particle_1995} after a comparison with identical objective and
peaks and equal numbers of objective evaluations
(Appendix~\ref{app:optimizer}). On six synthetic T4
lysozyme patterns and the measured T4 lysozyme pattern, at the largest tested
budget, which uses the default population size, it found the correct orientation in 94\% of single searches,
against 83\% for the best of four particle-swarm settings and 34--60\% for the
others, with one parameter setting for all test problems.

The workers can be organized in two ways. In our implementation the objective
is a method of the optimizer object, so SciPy's built-in pool sends it with
every task, and with it the optimizer object and its candidate tables, which are copied to the workers again in every generation. In
the persistent pool used here, the workers are started once, before the search,
and each keeps its own copy of the prepared $k$-d tree (step 3 of Algorithm~1). In every generation they receive only the trial orientations, three numbers per
member of the population, and return the objective values (steps 8--10).

\medskip
\noindent\rule{\columnwidth}{0.4pt}\par\noindent
\textbf{Algorithm 1.} Orientation search for a known cell and wavelength
band.\par\vspace{-2pt}\noindent\rule{\columnwidth}{0.4pt}
{\small
\begin{algorithmic}[1]
\Require peak vectors $\boldsymbol{q}_i$ and heights, cell $\boldsymbol{B}$,
band $[\lambda_\text{min},\lambda_\text{max}]$, tolerances $\tau_s=1^\circ$,
$\tau=0.35^\circ$
\State find the candidate shell $\mathcal{S}_i$ of each peak
(Eq.~\ref{eq:shell}); keep informative peaks, $p_i\le0.7$ (Eq.~\ref{eq:chance} with $N_i$)
\State select the search subset (up to 80 peaks); build the labeled $k$-d
tree $T$ of the points $(\hat{\boldsymbol{g}},\,i\Delta)$ of its candidates
\State start $W$ worker processes, each holding a copy of $T$ \Comment{once}
\For{restart $r=1,2,3$}
  \State initialize a population of 1024 points in $[0,1]^3$ (Sobol)
  \For{generation $=1,\dots,1000$}
    \State form trial points $\boldsymbol{x}$ by mutation and crossover
    \State send one chunk of trial points to each worker \Comment{3 numbers
    each}
    \State \textbf{on each worker:}
    $\boldsymbol{U}\gets\boldsymbol{U}(\boldsymbol{x})$; $\delta_i\gets$
    nearest neighbor of
    $(\boldsymbol{U}^\mathsf{T}\boldsymbol{R}^\mathsf{T}\hat{\boldsymbol{q}}_i,
    \,i\Delta)$
    in $T$; $C(\boldsymbol{x})$ by Eq.~(\ref{eq:cost}) at $\tau_s$
    \State gather $C$; keep the better of each trial point and its parent
  \EndFor
  \State assign $\boldsymbol{h}_i$ and refine $\boldsymbol{U}$
  (Eq.~\ref{eq:wahba}), reducing the tolerance from $\tau_s$ to $\tau$
  \State compute $p$ (Eq.~\ref{eq:pvalue}) at $\tau$ and at $\tau_s$ and keep
  the more significant; \textbf{if} $\log_{10}p<-5$ \textbf{then} stop
\EndFor
\State keep the most significant restart; \textbf{if} $\log_{10}p\ge-5$
\textbf{then} report the pattern as not indexed
\State index all peaks with its $\boldsymbol{U}$
\State \Return $\boldsymbol{U}\boldsymbol{B}$, $\boldsymbol{h}_i$, $\lambda_i$
\end{algorithmic}
}
\noindent\rule{\columnwidth}{0.4pt}
\medskip

\subsection{Assignment and orientation refinement}
\label{sec:refine}

After the search, Miller indices are assigned by Eq.~(\ref{eq:delta}) and
$\boldsymbol{U}$ is updated by the weighted alignment of observed and predicted
directions,
\begin{equation}
\boldsymbol{U}=\underset{\boldsymbol{U}\in
SO(3)}{\arg\min}\sum_{i:\,\delta_i\le\tau}w_i\left|\boldsymbol{R}^\mathsf{T}
\hat{\boldsymbol{q}}_i-\boldsymbol{U}\frac{\boldsymbol{B}\boldsymbol{h}_i}
{|\boldsymbol{B}\boldsymbol{h}_i|}\right|^2,
\label{eq:wahba}
\end{equation}
Wahba's problem, solved in closed form by singular value decomposition
\cite{wahba_least_1965,kabsch_solution_1976}. The same orthogonal Procrustes
problem updates the orientation of each image in Laue-DIALS
\cite{hewitt_laue-dials_2024}. Assignment and alignment
alternate, first on the search subset and then on all informative peaks, as
$\tau$ is reduced from $1^\circ$ to $0.35^\circ$.

\subsection{Chance-match score}
\label{sec:significance}

For a large cell at a wide band, many peaks are matched by chance whatever the orientation. At $\tau=0.35^\circ$, a random orientation of T4 lysozyme indexes
about two thirds of the observed peaks. The indexed fraction alone therefore
cannot show whether a solution is correct, and we judge a solution instead by
how far its number of indexed peaks exceeds the number expected for a random
orientation. This plays the role that the figures of merit $M_{20}$ and $F_N$
play in powder indexing, which compare the observed matches with the number
expected by chance for the proposed cell
\cite{de_wolff_simplified_1968,smith_fn_1979}. In orientation indexing, figures
of merit commonly count or fit the indexed reflections, or compare the best
solution with the second best \cite{morawiec_indexing_2022}; with $\sim10^4$
candidates per peak, the number of matches expected by chance is large enough
here that we compute it explicitly.

A cap of angular radius $\tau$ covers a fraction $(1-\cos\tau)/2$ of the
sphere. Under a random orientation each of the $n_i$ candidate directions of
peak $i$ is uniformly distributed, and treating them as independent, the
probability that at least one falls within $\tau$ of the peak is
\begin{equation}
p_i(\tau)=1-\left(\frac{1+\cos\tau}{2}\right)^{n_i}
\approx1-\exp\left(-\frac{n_i\tau^2}{4}\right).
\label{eq:chance}
\end{equation}
The expected number of peaks indexed by a random orientation is
$\Lambda(\tau)=\sum_ip_i(\tau)$. The count $X$ itself is a sum of Bernoulli
variables with probabilities $p_i$, which we approximate by a Poisson
distribution with mean $\Lambda$. If a solution indexes $K$ peaks, its score
is the Poisson tail
\begin{equation}
p=P(X\ge K)=\sum_{k\ge K}\frac{\Lambda^k e^{-\Lambda}}{k!},
\label{eq:pvalue}
\end{equation}
reported as $\log_{10}p$, with more negative values indicating a larger excess over
the chance background. To compare restarts and decide acceptance, a solution is
scored at both the final and the search tolerance, each with its own $\Lambda$,
and the more significant value is used (Section~\ref{sec:search}), since
residuals comparable to the final tolerance can place true matches just
outside it. The same per-peak probabilities select the informative peaks used
in the search.

Two approximations make $\log_{10}p$ a score for comparing restarts and
stopping the search, rather than a calibrated probability that a solution is
wrong. First, the candidate directions belong to one lattice and rotate
together, so Eq.~(\ref{eq:chance}) only approximates each $p_i$, and hence $\Lambda$. The matches of different peaks are also correlated, which affects
the shape of the distribution and in particular its tail. Second,
Eq.~(\ref{eq:pvalue}) refers to one orientation fixed in advance, whereas the
search examines many. Orientation space contains about $10^8$ distinguishable
orientations at $\tau=0.35^\circ$ (Appendix~\ref{app:chance}). A heuristic
correction for this search volume would therefore require $p$ to be much
smaller than $10^{-8}$ before accidental matches anywhere in orientation space
become unlikely, and for an adaptive search even this is only indicative. The acceptance threshold,
$\log_{10}p<-5$, is therefore an empirically tested stopping criterion, not a
false-positive rate of $10^{-5}$. Every final accepted solution reported below reaches
$\log_{10}p<-12$.

Two tests check the score. The first checks the reference model. For the 351
informative T4 lysozyme peaks (Section~\ref{sec:t4l}), $10^5$ random
orientations index on average 140.5 peaks within $0.35^\circ$, with a standard
deviation of 9.4, against $\Lambda=132.3$ and a Poisson standard deviation of
11.5. The model underestimates the mean by about 6\% but overestimates the
spread. A normal approximation using the measured mean and variance suggests
that the score is conservative near the acceptance threshold
(Appendix~\ref{app:chance}). The second test
checks the search itself, which could find chance solutions that random
sampling would miss. Each observed peak was rotated about the incident beam by
an independent random angle, which keeps its scattering angle and candidate
shell but removes the lattice. Each scrambled pattern was searched with the
full procedure of Section~\ref{sec:search}, with up to three restarts scored by
the more significant of the values at $\tau$ and $\tau_s$. None of 20 such null
searches for T4 lysozyme and 10 for mesolite was accepted, and the best scores were
$\log_{10}p=-3.2$ and $-3.4$. In half of the mesolite searches the score at
$\tau_s$ was the more significant, but none came closer to the threshold than
the best score at $\tau$. Thirty searches (90 restarts) cannot establish
a small false-positive rate, but none produced an accepted solution.

\subsection{Lattice refinement and centering}
\label{sec:centering}

With the assignments fixed, the lattice constants (constrained by the lattice
system) and $\boldsymbol{U}$ are refined by least squares on the vector
residuals
\begin{equation}
\boldsymbol{r}_i=\lambda_i\boldsymbol{R}\boldsymbol{U}\boldsymbol{B}
\boldsymbol{h}_i-\boldsymbol{q}_i,\qquad
\lambda_i=\frac{|\boldsymbol{q}_i|}{|\boldsymbol{B}\boldsymbol{h}_i|}\
\text{clipped to the band}.
\label{eq:refine}
\end{equation}
Since the scale is undetermined, one length is held fixed and only ratios and
angles are refined.
Centered cells are searched in a primitive cell, which has fewer systematically
absent candidates. Let the columns of $\boldsymbol{A}_C$ and $\boldsymbol{A}_P$
be the conventional and primitive basis vectors, with
$\boldsymbol{A}_P=\boldsymbol{A}_C\boldsymbol{P}$ for a centering matrix
$\boldsymbol{P}$ (Appendix~\ref{app:centering}). The metric tensors
$\boldsymbol{G}_C=\boldsymbol{A}_C^\mathsf{T}\boldsymbol{A}_C$ and
$\boldsymbol{G}_P$ are related by
$\boldsymbol{G}_P=\boldsymbol{P}^\mathsf{T}\boldsymbol{G}_C\boldsymbol{P}$ and
the indices transform as
$\boldsymbol{h}_C=\boldsymbol{P}^{-\mathsf{T}}\boldsymbol{h}_P$. The solution
is returned in the conventional setting as
$(\boldsymbol{U}\boldsymbol{B})_C=(\boldsymbol{U}\boldsymbol{B})
_P\boldsymbol{P}^\mathsf{T}$,
and its Miller indices obey the reflection conditions of the centering.

\subsection{Detector geometry calibration}
\label{sec:calibration}

Angular residuals of a few tenths of a degree limit indexing of large cells, so
the detector geometry of IMAGINE-X was calibrated with a garnet crystal
(Yb$_3$Al$_5$O$_{12}$, $Ia\bar{3}d$, $a=11.9386$~\AA) measured in 68
orientations with the 50 cameras installed at the time (24 in the back and 26
in the forward array). The 2722 reflections ($2\le\lambda\le10$~\AA) were
refined together with the goniometer axis, the reference orientation and an
angle offset for each run. Each array was treated as rigid, with the sample at
the origin. A common rotation of both arrays about the vertical axis is
degenerate with the goniometer, so only their relative rotation was refined.
Terms were added one at a time and kept only if they increased the fraction of
peaks indexed within $0.35^\circ$ in a held-out half of the runs, for which only
the angle offsets were refitted (Table~\ref{tab:calibration}). The relative
rotation and the vertical and horizontal positions of the arrays raised this
fraction from 25\% to 59\%. In-plane shifts of the individual cameras improved
the fitted half but not the held-out half, which indicates overfitting, and
were not retained.

\begin{table}
	\caption{Calibration of the IMAGINE-X detector geometry on garnet as terms are
	added to the model, with the parameters added, the fraction of peaks indexed within
	$0.35^\circ$ in the half of the runs used for the fit (training) and in the
	other half (held out, orientations refitted), and the median angular residual of
	the held-out reflections indexed within $0.35^\circ$ (1361 peaks in each
	half), which can rise as more marginal peaks are indexed.}
	\label{tab:calibration}
\centering
	\begin{tabular}{lcccc}
		Refined terms & Added & Training & Held out & Residual \\
		\hline\hline
		Goniometer$^a$ & -- & 32\% & 25\% & $0.22^\circ$ \\
		+ relative array rotation & 1 & 57\% & 33\% & $0.23^\circ$ \\
		+ array vertical positions & 2 & 75\% & 42\% & $0.24^\circ$ \\
		+ array horizontal positions & 4 & 90\% & 59\% & $0.21^\circ$ \\
		+ camera in-plane shifts & 100 & 95\% & 60\% & $0.18^\circ$ \\
		\hline
	\end{tabular}
	\par\smallskip
	{\footnotesize $^a$ Goniometer axis, reference orientation and an angle offset
	for each run.}
\end{table}

The second array is rotated by $-1.17(7)^\circ$ about the vertical axis
relative to the first, the arrays are displaced vertically by $6.3(2)$ and
$5.2(1)$~mm and horizontally by $(-1.5,-3.8)$ and $(4.0,2.2)$~mm (uncertainties
$\le0.5$~mm), and the goniometer axis is tilted by $0.44^\circ$ from the
vertical. Since the arrays can be rotated about the sample, these corrections
reflect the positions at which they are locked, together with the beam center.
A repeated calibration, from which Table~\ref{tab:calibration} is taken,
reproduced these values. The uncertainties are formal. Residuals within a run
are correlated, so they underestimate the true uncertainties, and the relative
rotation varied between $-0.76^\circ$ and $-1.17^\circ$ depending on which
other terms were refined. On the calibration runs, the median angular residual
fell from $0.30^\circ$ to $0.20^\circ$. The result is stored as ISAW DetCal
files.

The calibration was tested on a crystal it was not fitted to, 79 mesolite runs
($Fdd2$, $a=18.40$, $b=56.65$, $c=6.54$~\AA) measured on IMAGINE-X and refined
with the geometry held fixed. The fraction of held-out peaks indexed within
$0.35^\circ$ rose from 6\% with the nominal geometry to 38\%, and the refined
ratios $b/a=3.0773(15)$ and $c/a=0.3549(2)$ deviate from the reference values
(3.0788 and 0.3554) by 0.05\% and 0.14\%, against 0.21\% and 0.68\% before
calibration.

\subsection{Peak profiles}
\label{sec:profiles}

For display and validation (Figs.~\ref{fig:imagine} and \ref{fig:mandi}), each
indexed or predicted peak is fitted in a box around its position with a
two-dimensional Gaussian on a constant background,
\begin{equation}
I(\boldsymbol{x})=A\exp\left[-\tfrac{1}{2}(\boldsymbol{x}-\boldsymbol{\mu}
)^\mathsf{T}\boldsymbol{\Sigma}^{-1}(\boldsymbol{x}-\boldsymbol{\mu})\right]+
I_\text{bg},
\label{eq:ellipse}
\end{equation}
where $\boldsymbol{x}$ is the pixel position, $A$ the peak amplitude,
$\boldsymbol{\mu}$ the peak center, $I_\text{bg}$ the background, and the
covariance $\boldsymbol{\Sigma}$ has principal widths $\sigma_1$, $\sigma_2$
and orientation $\vartheta_\Sigma$. The parameters are obtained by least
squares with a robust (soft-$L_1$) loss, the integrated intensity is $2\pi
A\sigma_1\sigma_2$ with its uncertainty propagated from the parameter
covariance, and the peaks are drawn as the ellipses at three standard
deviations. The indexing itself uses only peak positions and, as weights, their
heights. The fitted profiles could, however, serve as an outlier diagnostic. Spurious detections such as hot pixels, streaks or edge artifacts tend to have
a low signal-to-noise ratio $I/\sigma(I)$ or implausible shapes (widths at the
fit bounds or extreme elongation), so they could be removed or down-weighted
before the search.

\section{Results} \bigskip

The algorithm was tested on three crystals spanning unit-cell volumes from
$1.7\times10^3$ to $3.1\times10^5$~\AA$^3$, measured on three instruments
(Table~\ref{tab:accuracy}). For every dataset, the cell and wavelength band
were supplied, with no initial orientation. All timings are wall times on an
Intel i9-13950HX workstation with 16 worker processes and the default search
settings of Section~\ref{sec:search} (1024 population members, at most 1000
generations, up to three restarts).

\begin{table*}
	\TABtrue\refstepcounter{table}
	\caption{Indexing results. Found: peaks detected. Informative: peaks with
	$p_i\le0.7$, used for refinement and scoring, from which the search subset
	of up to 80 peaks is drawn (Section~\ref{sec:search}). Indexed: share of
	informative peaks indexed
	within $\tau=0.35^\circ$, with the share expected for a random orientation
	(chance) and counts in parentheses, with $p$ from Eq.~(\ref{eq:pvalue}). Residual:
	median angular error of indexed reflections. Time: wall time of the
	orientation search with 16 workers, for the run shown (mesolite, T4
	lysozyme) or the median over the 93 settings (garnet).}
	\label{tab:accuracy}
	\centering
	{\small\setlength{\tabcolsep}{2.5pt}
	\begin{tabular}{llcccccccc}
		Crystal & Instrument & \shortstack{$\lambda$\\ (\AA)} & Found &
		Informative & Indexed & Chance & $\log_{10}p$ & Residual & Time \\[2pt]
		\hline\hline
		\shortstack[l]{Mesolite\\ ($Fdd2$)} & IMAGINE & 2--4.5 & 173 & 173 &
		\shortstack{91\%\\ (157)} & \shortstack{3.1\%\\ (5.3)} & $-167$ &
		$0.06^\circ$ & 43~s \\[6pt]
		\shortstack[l]{Garnet\\ ($Ia\bar{3}d$)} & MANDI & 1--3 & 43 & 43 &
		\shortstack{100\%\\ (43)} & \shortstack{4.2\%\\ (1.8)} & $-41.7$ &
		$0.04^\circ$ & 45~s \\[6pt]
		\shortstack[l]{T4 lysozyme\\ ($P3_221$)} & IMAGINE-X & 2--10 & 833 & 351 &
		\shortstack{65\%\\ (229)} & \shortstack{38\%\\ (132)} & $-13.8$ &
		$0.19^\circ$ & 101~s \\
		\hline
	\end{tabular}}
	\par\smallskip
	{\footnotesize Garnet: medians over 93 goniometer settings (3851 reflections), with 84.3\% of assignments identical to time-of-flight indexing, 15.5\% harmonics,
	none different.}
\end{table*}

\subsection{Mesolite on IMAGINE}

\begin{figure}
	\includegraphics[width=0.85\textwidth]{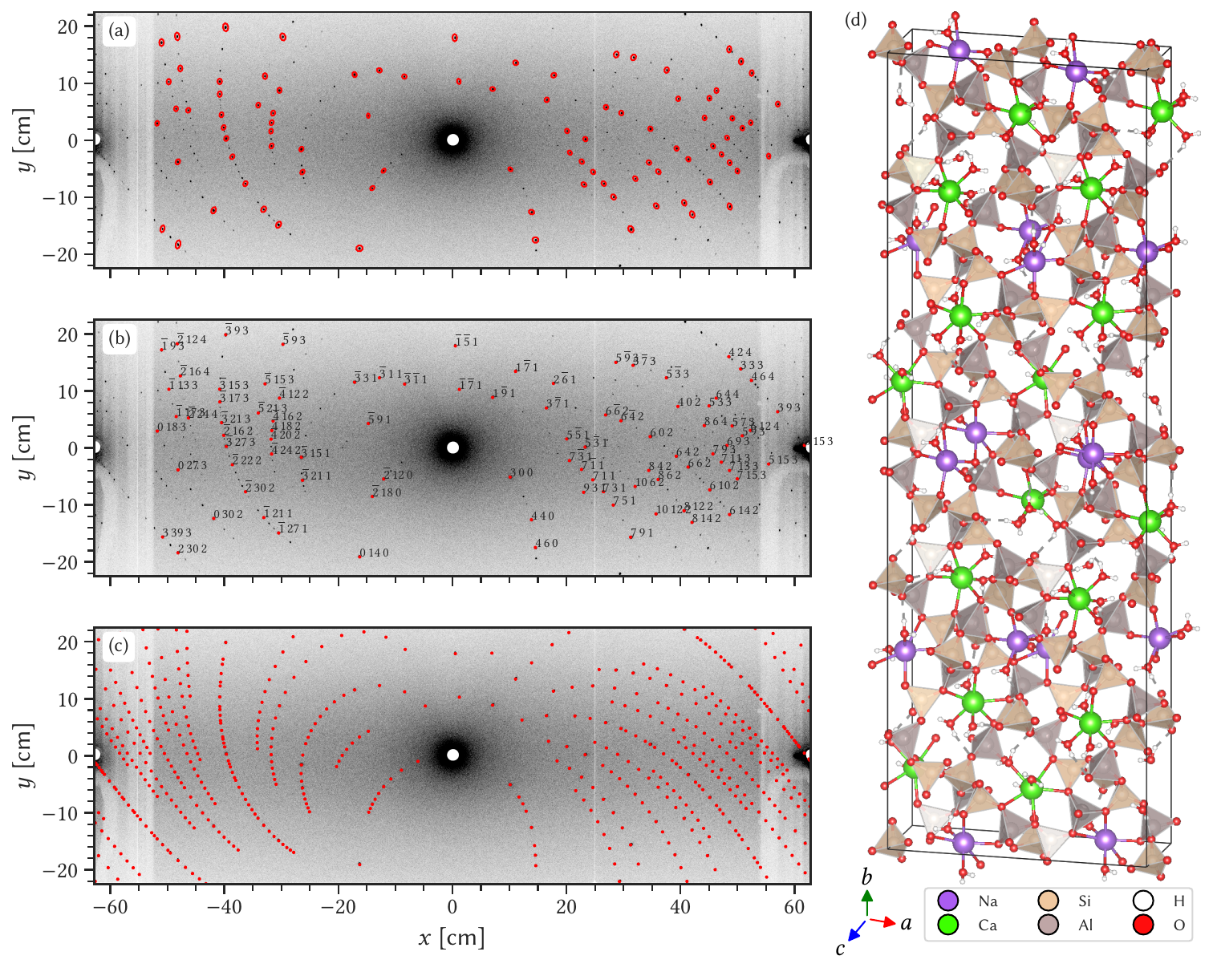}\\
	\caption{
		Mesolite (Na$_2$Ca$_2$Al$_6$Si$_9$O$_{30}\cdot8$H$_2$O, $Fdd2$, $a=18.39$,
		$b=56.55$, $c=6.54$~\AA) measured at IMAGINE on the cylindrical image plate
		in a single 15-minute exposure at $2\le\lambda\le4.5$~\AA, shown in unrolled
		detector coordinates.
		(a) Peaks found by the peak search, drawn as the $3\sigma$ ellipses of the
		fitted profiles (Section~\ref{sec:profiles}).
		(b) Peaks indexed with the determined $UB$ matrix, labeled with $hkl$.
		(c) Reflections predicted within the band down to $d_\text{min}=1$~\AA.
		(d) Crystal structure of mesolite with the long $b$ axis vertical
		\cite{artioli_x-ray_1986,ross_crystalline_1992}; crystal axes $a$, $b$ and
		$c$ in red, green and blue.
	}
	\label{fig:imagine}
\end{figure}

A single 15-minute image-plate exposure of mesolite ($Fdd2$, $a=18.39$,
$b=56.55$, $c=6.54$~\AA) at $2\le\lambda\le4.5$~\AA{} gave 173 peaks, of which
157 were indexed within $0.35^\circ$ against 5.3 expected for a random
orientation ($\log_{10}p=-167$), with a median angular residual of $0.06^\circ$
(Fig.~\ref{fig:imagine}). The long $b$ axis makes the reflections dense along
one direction, but the narrow band keeps the number of candidates per peak
small, so a single search of 43~s sufficed. Ten searches with independent
seeds found the identical orientation.

\subsection{Garnet on MANDI: comparison with time-of-flight indexing}

\begin{figure}
	\includegraphics[width=\textwidth]{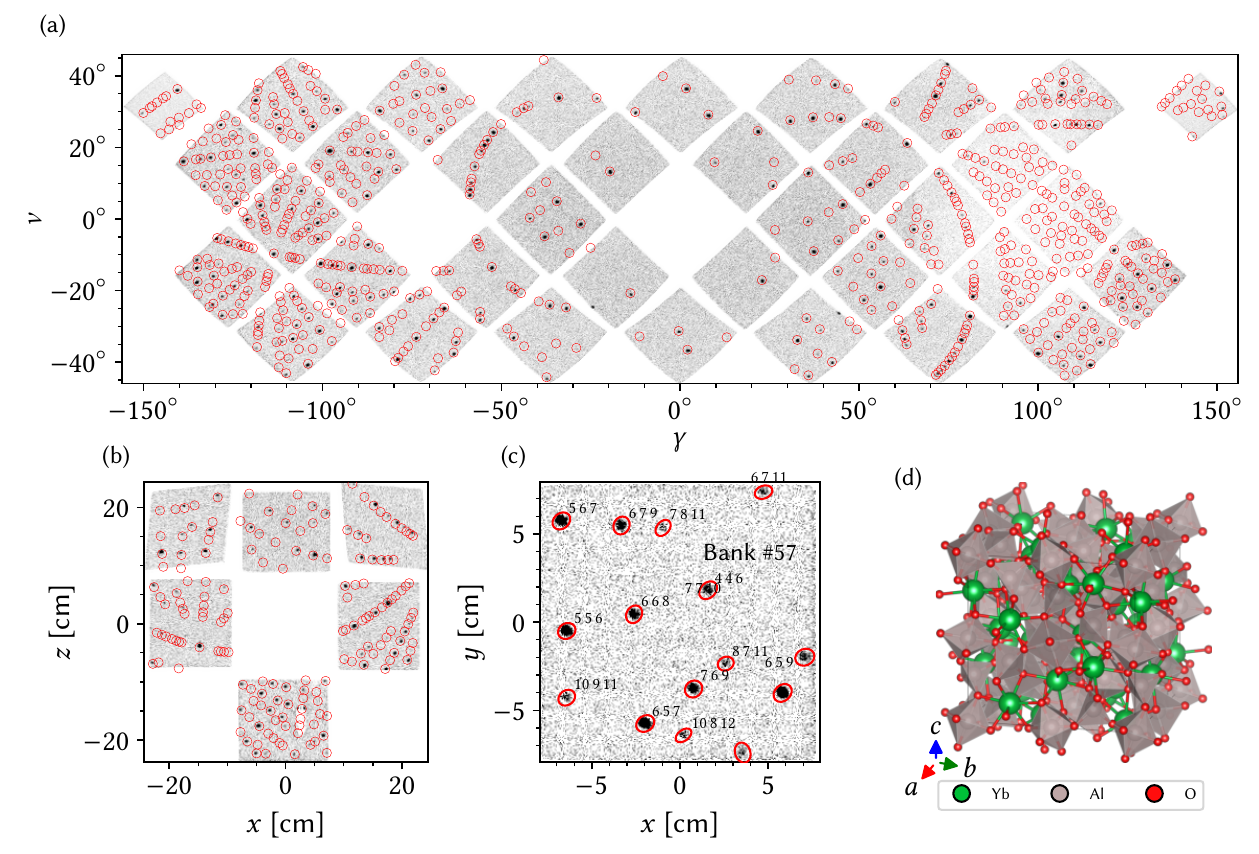}\\
	\caption{
		Garnet (Yb$_3$Al$_5$O$_{12}$, $Ia\bar{3}d$, $a=11.9386$~\AA) measured at
		MANDI (run 12607, 24~min, $1\le\lambda\le3$~\AA), with the time-of-flight
		information not used for indexing.
		(a) Detector counts of the main array of Anger cameras displayed with
		horizontal and vertical angles $\gamma$ and $\nu$, with reflection positions
		predicted down to $d_\text{min}=0.7$~\AA{} (red circles). The smaller detectors in the upper left and right corners are prototype silicon
		photomultiplier (SiPM) detectors.
		(b) The bottom panel of detectors, mounted differently from the main array,
		in physical units.
		(c) One camera (bank 57) with indexed reflections labeled with $hkl$ and
		drawn as the $3\sigma$ ellipses of the fitted profiles
		(Section~\ref{sec:profiles}).
		(d) Crystal structure of the garnet \cite{dobrzycki_structure_2004}; crystal
		axes $a$, $b$ and $c$ in red, green and blue.
	}
	\label{fig:mandi}
\end{figure}

MANDI \cite{coates_macromolecular_2015} resolves the wavelength of every
reflection by time-of-flight, which gives an independent reference. The 3851
reflections of a garnet calibration series (Yb$_3$Al$_5$O$_{12}$, $Ia\bar{3}d$,
$a=11.9386$~\AA, 93 goniometer settings, $1\le\lambda\le3$~\AA) were indexed by
the time-of-flight UB matrix determined with Mantid
\cite{arnold_mantiddata_2014}. We discarded their wavelengths, kept only their
directions, and indexed each setting independently as if it were a quasi-Laue
measurement (Fig.~\ref{fig:mandi}). All 93 searches succeeded, with
$\log_{10}p$ between $-75.8$ and $-17.6$, and 99.8\% of the reflections were
indexed with a median residual of $0.04^\circ$. The assignments agree with
time-of-flight for 84.3\% of the reflections, and the remaining 15.5\% are
harmonics ($n\boldsymbol{h}$ in place of $\boldsymbol{h}$), which diffract into
the same direction and cannot be distinguished without wavelength
\cite{cruickshank_multiplicity_1987}; no peak received a different
reciprocal-lattice row. With the band restricted to the observed
$1\le\lambda\le2$~\AA, the harmonic share falls to 3.2\% while the score
is unchanged, since the wider band adds only lower orders of reciprocal-lattice
rows already among the candidates. For each setting, the wavelengths implied by the assignment,
$\lambda_i=|\boldsymbol{q}_i|/|\boldsymbol{B}\boldsymbol{h}_i|$, were compared
with the time-of-flight values over all indexed reflections. The median across
the 93 settings of the within-setting median absolute relative difference is
0.07\%. Harmonic assignments are included. Their wavelengths differ by an
integer factor, but they form a minority of the observations, so the median
primarily characterizes the non-harmonic assignments. The orientations differ from the
common time-of-flight UB by a median of $0.27^\circ$, which reflects the
per-setting goniometer errors absorbed by the independent searches rather than
errors of the indexing.

The time-of-flight peak list contains no false detections, so we also tested
the effect of spurious peaks. For 19 of the settings, directions of peaks from
other settings, which fall on the detector but are unrelated to the lattice in
that setting, were added in numbers equal to 10, 25, 50 and 100\% of the true
peaks. Every search still found the correct orientation and indexed at least 97\%
of the true peaks, with a median $\log_{10}p$ of $-32$ and a worst case of $-12.8$ even
with as many spurious as true peaks. The spurious peaks were indexed at the
rate expected by chance (4.3\% against 4.1\% when they equaled the true peaks
in number), so they weaken the score but do not mislead the search.

\subsection{T4 lysozyme on IMAGINE-X}
\label{sec:t4l}

\begin{figure}
\includegraphics[width=\textwidth]{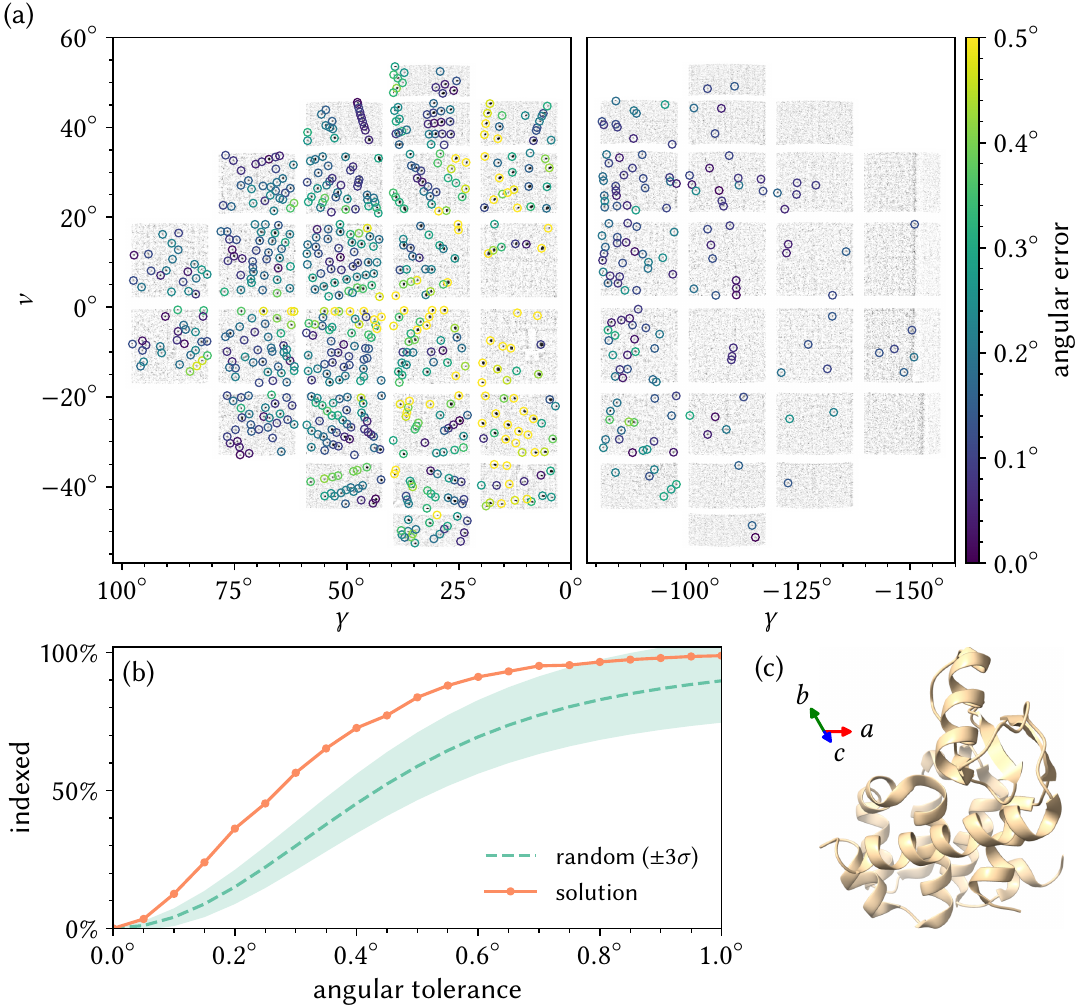}\\
\caption{
T4 lysozyme ($P3_221$, refined room-temperature cell $a=61.1$, $c=96.8$~\AA)
measured at IMAGINE-X (run 1816, $2\le\lambda\le10$~\AA) with the 50 cameras
installed at the time, after array-level calibration of the detector geometry
on garnet (Section~\ref{sec:calibration}).
(a) Background-subtracted counts of the forward (left) and back (right) arrays
displayed with horizontal and vertical angles $\gamma$ and $\nu$, with the 833
peaks found by the matched-filter search colored by their angular error to the
nearest candidate reflection direction within the band (values above
$0.5^\circ$ in the top color). At high scattering angles the candidate
directions are dense, so small errors there are also expected by chance.
(b) Percentage of the 351 informative peaks (chance-match probability below
0.7, used for refinement and scoring) indexed by the refined solution as a
function of angular tolerance, compared with the expectation (dashed) and
$\pm3\sigma$ Poisson band (shaded) for a random orientation.
(c) Crystal structure of T4 lysozyme \cite{li_neutron_2017}; crystal axes $a$,
$b$ and $c$ in red, green and blue.
}
\label{fig:imaginex}
\end{figure}

T4 lysozyme ($P3_221$, $V=3.1\times10^5$~\AA$^3$) at the full band of IMAGINE-X
($2\le\lambda\le10$~\AA) is the hardest case. A high-angle peak has up to
$\sim10^4$ candidate reflections, and a random orientation indexes about two
thirds of the found peaks within $0.35^\circ$. With the matched-filter peak search,
833 peaks were found, of which 351 were informative (Section~\ref{sec:search};
Fig.~\ref{fig:imaginex}). The solution indexes 229 of them within $0.35^\circ$
against 132 expected by chance ($\log_{10}p=-13.8$), and 127 within $0.2^\circ$
against 53 ($\log_{10}p=-17.4$), and ten searches with independent seeds
converged to the same orientation within $0.13^\circ$ (median 75~s each,
against 101~s for the run in Table~\ref{tab:accuracy}).

The solution was tested on the five following runs of the same crystal,
each rotated by a further $10^\circ$ about the vertical
goniometer axis (Table~\ref{tab:series}). Each run was indexed from the refined
orientation of the previous run, rotated by the goniometer step, and refined
locally. This tests the transfer of an established orientation and its local
refinement, not independent global indexing of each pattern. Successful
prediction across the series provides evidence that the orientation of the first run
is correct. The predicted orientations already passed the threshold before
refinement ($\log_{10}p$
between $-7.2$ and $-11.0$), refinement moved them by only
$0.18$--$0.32^\circ$, and the refined solutions reached $\log_{10}p$ between
$-12.9$ and $-16.5$. A rotation of the same size in the opposite sense did not
index the second run ($\log_{10}p=-1.7$). Over the series, the refined orientations
rotate by $49.5^\circ$ about an axis $1.1^\circ$ from vertical, against
$50^\circ$ expected, with $c/a$ between 1.588 and 1.601. The third run
($\omega=20^\circ$) was also indexed on its own by the global search. Five searches with independent seeds
converged within $0.3^\circ$ of the orientation predicted by the series and
reached $\log_{10}p\approx-18.6$ after refinement.

\begin{table}
\caption{Rotation series of T4 lysozyme on IMAGINE-X. Run 1816 was indexed by
the global search (seed 1, which gives a slightly more significant solution
than the run of Table~\ref{tab:accuracy}). Every run starts from the same
room-temperature cell ($a=61.1$, $c=96.8$~\AA). Each later run was indexed from the refined orientation of
the previous run rotated by $10^\circ$ about the vertical goniometer axis
(predicted) and then refined locally (refined). Indexed: informative peaks
indexed within $0.35^\circ$ against the number expected for a random
orientation. The local refinement also refines the cell ratio $c/a$, which
changes the candidate shells and hence the expected count. $\omega$: nominal
goniometer rotation relative to run 1816. Correction: rotation applied by the
local refinement.}
\label{tab:series}
\centering
{\small
\begin{tabular}{lccccccc}
Run & $\omega$ & Informative & \multicolumn{2}{c}{Predicted} &
\multicolumn{2}{c}{Refined} & Correction \\
 & & peaks & Indexed & $\log_{10}p$ & Indexed & $\log_{10}p$ & \\
\hline\hline
1816 & $0^\circ$ & 351 & -- & -- & 237 / 132 & $-15.8$ & -- \\
1818 & $10^\circ$ & 383 & 225 / 142 & $-10.1$ & 251 / 142 & $-15.9$ & $0.32^\circ$ \\
1820 & $20^\circ$ & 379 & 218 / 143 & $-8.4$ & 243 / 143 & $-13.7$ & $0.18^\circ$ \\
1822 & $30^\circ$ & 357 & 197 / 131 & $-7.2$ & 225 / 132 & $-12.9$ & $0.32^\circ$ \\
1824 & $40^\circ$ & 347 & 215 / 131 & $-11.0$ & 239 / 132 & $-16.5$ & $0.24^\circ$ \\
1826 & $50^\circ$ & 347 & 198 / 129 & $-7.9$ & 224 / 130 & $-13.3$ & $0.32^\circ$ \\
\hline
\end{tabular}}
\end{table}

Indexing required the array-level calibration. With the nominal geometry the
best orientation did not pass the threshold ($\log_{10}p\approx-2$), within the range
of the null searches (Section~\ref{sec:significance}). This highlights the
importance of a well-calibrated instrument, since for a large cell at a wide
band the angular tolerance that separates true from chance matches is only a
few tenths of a degree. The search was first
run with the cell of the 80~K structure 5VNQ \cite{li_neutron_2017}
($c/a=1.559$). On an earlier list of 282 peaks from a local-maximum search, it
found an orientation of marginal significance ($\log_{10}p\approx-6$), and two
of four seeds did not pass the threshold. Searches with trial room-temperature
cells ($c/a=1.57$--1.59) converged to within $0.5^\circ$ of the same
orientation with more significant scores, and refining $c/a$ from one of these solutions,
with $a$ fixed, gave $c/a=1.585$. This room-temperature cell ($a=61.1$,
$c=96.8$~\AA), longer in $c$ than the 80~K cell, is used for all T4 lysozyme
results here. A plausible
refined cell is a consistency check on the solution, although not an
independent one, since the cell and the score both use the assigned
reflections. The
remaining angular errors are largest on the forward low-angle cameras
(Fig.~\ref{fig:imaginex}a), which points to per-camera geometry beyond the
array-level model. Many of the peaks outside the $0.35^\circ$ tolerance appear
to be true reflections with larger errors. Increasing the tolerance to $0.7^\circ$ indexes 95\% of the
informative peaks (334 of 351; Fig.~\ref{fig:imaginex}b), and the peaks that
need the larger tolerance are concentrated at small scattering angles
($2\theta<35^\circ$ for 47\% of them, against 15\% of the peaks indexed within
$0.35^\circ$) and near the edges of the cameras (40\% within 16 of the 128
pixels of an edge, against 21\%). In our experience
with time-of-flight diffraction, these are also the regions where the detector
geometry is hardest to calibrate, which suggests that much of the remaining
error comes from the calibration. More chance matches are also expected at the larger tolerance
(Fig.~\ref{fig:imaginex}b), so not every additional match at $0.7^\circ$ need
be correct.

\subsection{Computational performance}

\begin{figure}
\includegraphics[width=\columnwidth]{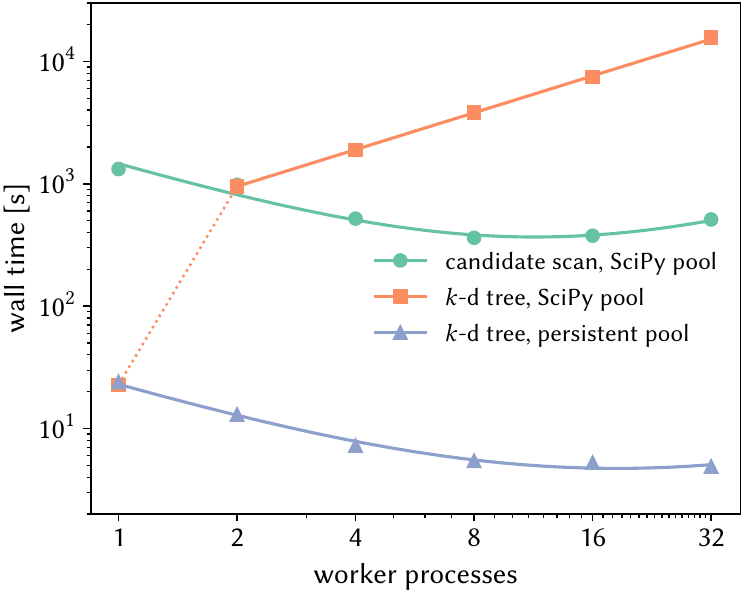}\\
\caption{
Wall time of one differential-evolution search for T4 lysozyme (run 1816; 80
peaks, 60 population members per parameter, 300 generations, same seed) against the number of worker
processes (Intel i9-13950HX, 32 threads), for the scan over all candidate
reflections and the $k$-d tree, each distributed with SciPy's pool, and for the
$k$-d tree distributed with a persistent pool. Lines: fits of
Eq.~(\ref{eq:amdahl}), with $T_s=137\pm68$~s, $T_p=1312\pm182$~s,
$c=10.1\pm3.3$~s (scan) and $T_s=2.5\pm0.6$~s, $T_p=20.4\pm1.9$~s,
$c=0.06\pm0.03$~s (tree, persistent pool), and $c=476\pm13$~s with negligible
$T_s$ and $T_p$ (tree, SciPy pool, fitted from 2 workers, since SciPy evaluates
serially without a pool for one worker, and the dotted line marks this switch). All
searches reach the same orientation.
}
\label{fig:timing}
\end{figure}

The cost of a trial orientation is dominated by the nearest-candidate search of
Eq.~(\ref{eq:delta}). Fig.~\ref{fig:timing} compares complete searches for T4
lysozyme in a reduced configuration, smaller than the default search (the pattern of
Table~\ref{tab:accuracy}, 80 search peaks, 60 population members per parameter, rounded up to 256, 300
generations, same seed), and every search reaches the same orientation. All three configurations use
SciPy's \texttt{differential\_evolution} \cite{virtanen_scipy_2020}; they
differ only in how the objective is evaluated and how the population is
distributed over worker processes (Table~\ref{tab:configurations}). Replacing
the scan over all candidates by the $k$-d tree shortens a serial search from
1320~s to 24~s. Distributing the population with SciPy's built-in pool re-sends
the optimizer, including its candidate tables, with every task
(Section~\ref{sec:search}), so the scan stops improving beyond 8 workers
(362~s) and slows down at 32 (511~s). The two changes are separated by the
third curve, the $k$-d tree distributed with SciPy's pool. Serially it takes
23~s, as with the persistent pool, but with workers its wall time grows in
proportion to their number, from 949~s with 2 workers to 15\,671~s with 32.
Sending the optimizer, now including the $k$-d tree, to the workers costs far
more than the search it distributes, so the fast objective pays off only
together with the persistent pool. This overhead arises because our objective
is a method of an optimizer object that carries the candidate tables, so it is a property of this implementation rather than of SciPy's parallel evaluation in
general. A persistent pool that receives the $k$-d tree once and is sent only
trial orientations reaches 4.9~s with 32 workers, 270 times faster than the
serial scan (1320~s with one worker). The wall times follow

\begin{equation}
T(n)=T_s+\frac{T_p}{n}+c\,n,
\label{eq:amdahl}
\end{equation}
Amdahl's law with a serial part $T_s$ and a parallel part $T_p$, plus an
overhead $c$ per worker. Both implementations parallelize about 90\% of the
work ($T_p/(T_s+T_p)=0.91$ for the scan and 0.89 for the tree), and the remainder is the serial per-generation step of differential evolution. With SciPy's pool,
the per-worker overhead is $c=10\pm3$~s for the scan and $476\pm13$~s for the
$k$-d tree, whose larger optimizer is more expensive to send. The persistent pool reduces it to $0.06\pm0.03$~s, which moves the optimum $n=\sqrt{T_p/c}$
from about 11 workers for the scan to 18 for the tree.

\begin{table}
\caption{Configurations compared in Fig.~\ref{fig:timing}. All use SciPy's
differential evolution, and \texttt{workers} is its argument for evaluating the
population in parallel.}
\label{tab:configurations}
\begin{tabular}{lll}
Configuration & Objective & Parallel evaluation \\
\hline\hline
Candidate scan, SciPy pool & full scan & SciPy, \texttt{workers=}$n$ \\
$k$-d tree, SciPy pool & $k$-d tree & SciPy, \texttt{workers=}$n$ \\
$k$-d tree, persistent pool & $k$-d tree & persistent pool \\
\hline
\end{tabular}
\end{table}

\section{Discussion} \bigskip

The chance-match score changes how a quasi-Laue indexing result can be judged.
For small cells, such as mesolite and garnet, a random orientation indexes a
few percent of the peaks, and the indexed fraction is a sufficient criterion.
For T4 lysozyme at the full band of IMAGINE-X, a random orientation indexes
almost 40\% of the informative peaks and about two thirds of all peaks. A correct and an incorrect orientation then differ by tens of peaks out of
hundreds, and the comparison with the chance expectation gives an explicit
measure of their separation. The
same comparison identifies the peaks to use in the search. Excluding the 58\%
of peaks that a random orientation matches more than 70\% of the time reduced
$p$ of the T4 lysozyme search solution by more than five orders of magnitude,
with $\log_{10}p$ decreasing from $-7.9$ to $-13.3$.

The method has limitations. The unit cell must be known, and because
wavelengths are not measured, only ratios of cell lengths and the angles can be refined. The absolute scale must come from elsewhere. The repeated T4 lysozyme
searches test orientation recovery for the prepared room-temperature cell. They do not establish unattended recovery from the 80~K reference cell, which gave
only a marginal solution, so for large cells an approximate cell may first
have to be refined. On MANDI, 15.5\% of the
reflections were assigned a different harmonic order than by time-of-flight at
a band of 1--3~\AA, and 3.2\% at 1--2~\AA. Deconvolution of harmonic overlaps
is left to integration
\cite{cruickshank_multiplicity_1987,campbell_lauegen_1995}. For large cells,
the angular precision required is a few tenths of a degree, and the T4 lysozyme
data could be indexed only after the detector geometry had been calibrated. The remaining errors on the low-angle cameras suggest that further calibration of
the individual cameras could improve it. Finally, each image is assumed to
contain one lattice.

This work was developed alongside subhkl
(\url{https://github.com/neutrons/subhkl}), an open-source package for the
reduction of IMAGINE-X data under development by several of the authors. subhkl
uses a related global search by differential evolution, implemented for
graphics processors, together with refinement of the instrument geometry,
integration of overlapping and harmonic reflections, wavelength normalization
and scaling, and will be described separately. The chance-match score, the
selection of informative peaks and the $k$-d tree search introduced here are
complementary to that implementation and can be adopted within it. Interactive
indexing, with visualization of the solutions, may be made available in
NeuXtalViz \cite{morgan_neuxtalviz_2026}.

\section{Conclusion} \bigskip

We have described an algorithm that indexes quasi-Laue neutron diffraction
patterns for a known cell and wavelength band without measuring wavelengths. A
global search by differential evolution, an exact $k$-d tree search of
candidate reflection directions and a chance-match score against random
orientations together make the indexing automatic, fast and verifiable, from
small inorganic cells to protein crystals at a wide band. On MANDI, the
indexing agrees with time-of-flight for every indexed reflection up to the
harmonic order, and on the new IMAGINE-X instrument it indexes T4 lysozyme over
the full 2--10~\AA{} band, where ten independent searches converge to the same
orientation within $0.13^\circ$. Rotated by the goniometer and refined
locally, that orientation also indexes five further runs of a $50^\circ$
rotation series.

The implementation described here uses Mantid \cite{arnold_mantiddata_2014} for the instrument
geometry, data loading and peak prediction, and SciPy
\cite{virtanen_scipy_2020} for the optimization. The source code, together with
the scripts that reproduce the results, is available at
\url{https://github.com/zjmorgan/quasi-laue-reduction}.

\appendix
\section{Rotation matrix}
\label{app:rotation}

The orientation search of Section~\ref{sec:orientation} does not vary the nine
elements of $\boldsymbol{U}$ directly, which would also produce matrices that
are not rotations, but the three angles of an axis and a rotation about it.
Every set of angles then gives a proper rotation, and the search space has the
three dimensions of orientation space. For reference, the rotation matrix of
Eq.~(\ref{eq:rodrigues}) is written out here in terms of these angles.

With $\hat{\boldsymbol{n}}=(n_x,n_y,n_z)$ and $c=\cos\omega$, $s=\sin\omega$,
Eq.~(\ref{eq:rodrigues}) is
\begin{equation}
\boldsymbol{U}=
\begin{bmatrix}
c+n_x^2(1-c) & n_xn_y(1-c)-n_zs & n_xn_z(1-c)+n_ys\\
n_yn_x(1-c)+n_zs & c+n_y^2(1-c) & n_yn_z(1-c)-n_xs\\
n_zn_x(1-c)-n_ys & n_zn_y(1-c)+n_xs & c+n_z^2(1-c)
\end{bmatrix},
\label{eq:rotation_matrix}
\end{equation}
with $n_x=\sin\vartheta\cos\varphi$, $n_y=\sin\vartheta\sin\varphi$,
$n_z=\cos\vartheta$, and $0\le\vartheta\le\pi$, $0\le\varphi<2\pi$,
$0\le\omega\le\pi$.

\section{Uniform sampling of orientations}
\label{app:sampling}

A global search is only unbiased if its trial orientations cover orientation
space evenly. Otherwise some orientations are examined far more often than
others, and a crystal in a sparsely sampled orientation may be missed. Drawing
the axis angles and the rotation angle uniformly does not achieve this, because
equal ranges of these angles do not correspond to equal volumes of orientation
space. This appendix shows how the parameterization of Eq.~(\ref{eq:uniform})
corrects for this, and illustrates the difference.

\begin{figure}
\includegraphics[width=\textwidth]{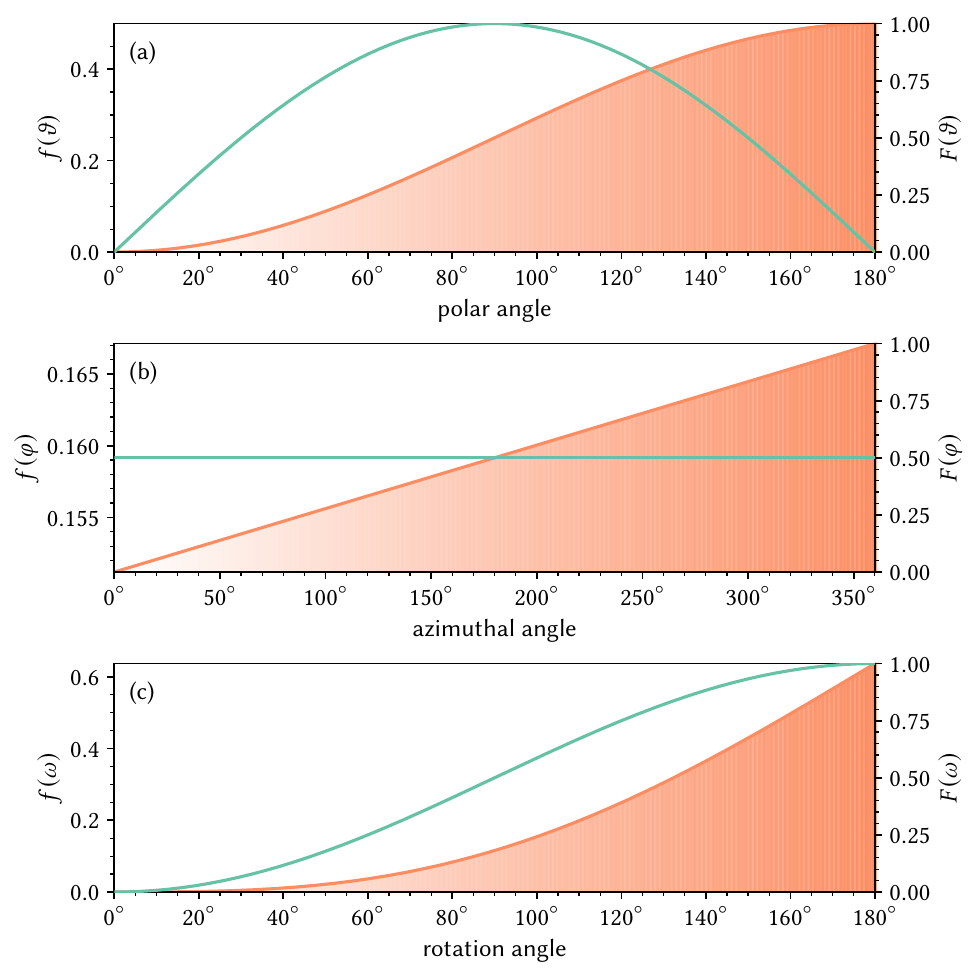}\\
\caption{
Probability densities $f$ (green, left axes) and cumulative distributions $F$
(orange, right axes) of the orientation parameters for uniformly random
rotations.
(a) Polar angle of the rotation axis, $f(\vartheta)=\tfrac{1}{2}\sin\vartheta$.
(b) Azimuthal angle of the rotation axis, $f(\varphi)=1/2\pi$.
(c) Rotation angle, $f(\omega)=(1-\cos\omega)/\pi$.
Drawing $u_0$, $u_1$ and $u_2$ uniformly and inverting $F$
(Eq.~\ref{eq:uniform}) samples each distribution.
}
\label{fig:probability}
\end{figure}

\begin{figure}
\includegraphics[width=\textwidth]{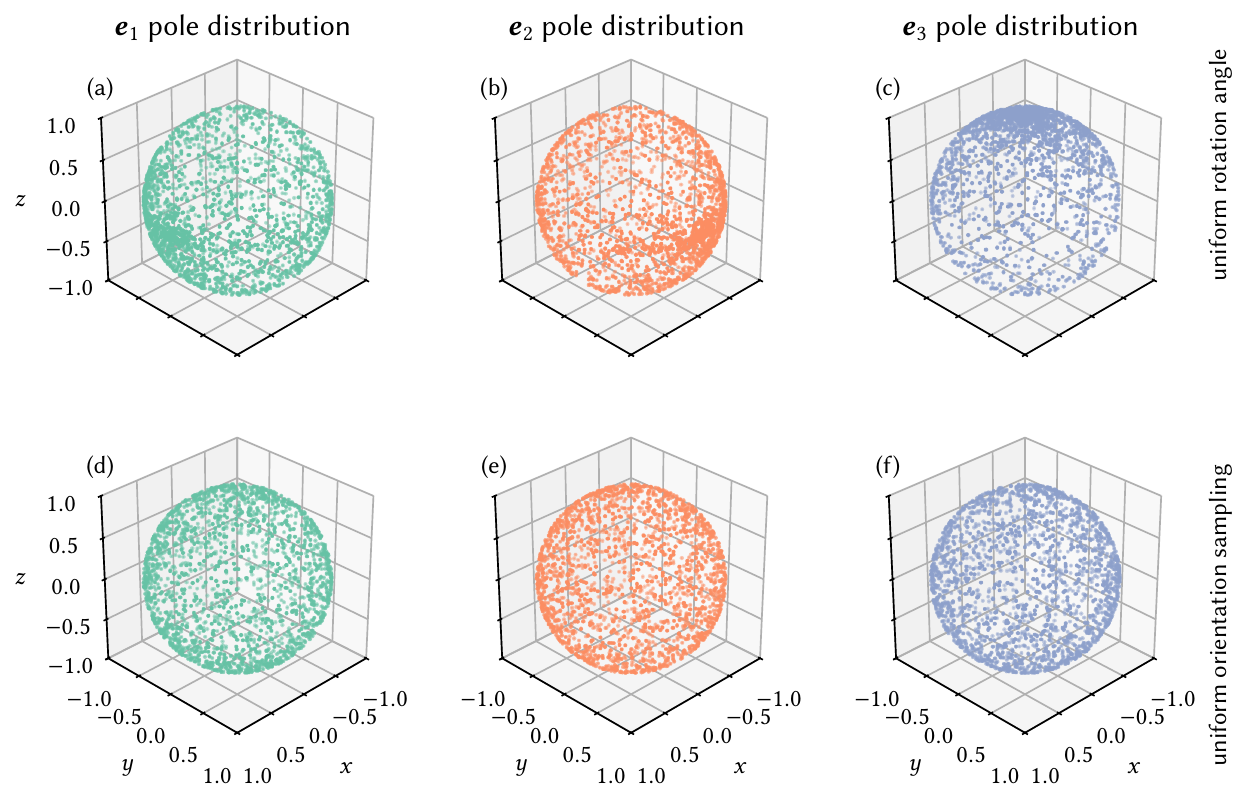}\\
\caption{
Directions of the rotated basis vectors $\boldsymbol{U}\boldsymbol{e}_1$,
$\boldsymbol{U}\boldsymbol{e}_2$ and $\boldsymbol{U}\boldsymbol{e}_3$ for
random orientations.
(a)--(c) Rotation axis drawn uniformly on the sphere (uniform sphere point
picking) but rotation angle $\omega$ drawn uniformly in $[0,\pi]$, which
over-weights small rotations and concentrates each pole near its starting
direction.
(d)--(f) Uniform orientation sampling by Eq.~(\ref{eq:uniform}), which covers
the sphere uniformly.
}
\label{fig:sampling}
\end{figure}

Uniform $(u_0,u_1,u_2)$ give $\cos\vartheta$ uniform on $[-1,1]$, $\varphi$
uniform on $[0,2\pi)$ and $\omega$ with cumulative distribution
$(\omega-\sin\omega)/\pi$, the angle distribution of uniformly random rotations
\cite{miles_random_1965}. Sampling $\vartheta$ or $\omega$ uniformly instead
would concentrate trial orientations near the poles of the axis sphere or near
the identity. The distributions of the three parameters are shown in
Fig.~\ref{fig:probability}, and the resulting orientations in
Fig.~\ref{fig:sampling}.

\section{Centering matrices}
\label{app:centering}

For a centered lattice, the conventional cell contains two, three or four
lattice points, and many of its reflections are systematically absent.
Searching in the conventional cell would include these absent reflections among
the candidates of every peak, adding chance matches without adding information.
The search is therefore carried out in a primitive cell of the same lattice
(Section~\ref{sec:centering}), and the solution is converted back to the
conventional setting. This appendix lists the matrices used for that
conversion.

The columns of $\boldsymbol{P}$ are the primitive basis vectors in units of the
conventional ones, $\boldsymbol{A}_P=\boldsymbol{A}_C\boldsymbol{P}$:
\begin{align}
\boldsymbol{P}_A&=\frac{1}{2}\begin{bmatrix}2&0&0\\0&1&-1\\0&1&1\end{bmatrix},
&
\boldsymbol{P}_B&=\frac{1}{2}\begin{bmatrix}1&0&-1\\0&2&0\\1&0&1\end{bmatrix},
\nonumber\\
\boldsymbol{P}_C&=\frac{1}{2}\begin{bmatrix}1&1&0\\-1&1&0\\0&0&2\end{bmatrix},
&
\boldsymbol{P}_I&=\frac{1}{2}\begin{bmatrix}-1&1&1\\1&-1&1\\1&1&-1\end{bmatrix}
,
\nonumber\\
\boldsymbol{P}_F&=\frac{1}{2}\begin{bmatrix}0&1&1\\1&0&1\\1&1&0\end{bmatrix}, &
\boldsymbol{P}_R&=\frac{1}{3}\begin{bmatrix}2&-1&-1\\1&1&-2\\1&1&1\end{bmatrix}
,
\label{eq:centering}
\end{align}
with $\boldsymbol{P}_R$ for the obverse setting of a rhombohedral lattice on
hexagonal axes. Each has a positive determinant, equal to the inverse of the
number of lattice points in the conventional cell ($\tfrac{1}{2}$,
$\tfrac{1}{4}$ or $\tfrac{1}{3}$), so the primitive cell has the same
handedness and $\boldsymbol{U}$ remains a proper rotation. Miller indices
transform as $\boldsymbol{h}_C=\boldsymbol{P}^{-\mathsf{T}}\boldsymbol{h}_P$,
which gives integer $\boldsymbol{h}_C$ obeying the reflection conditions of the
centering. The reflection conditions applied to the conventional indices are
listed in Table~\ref{tab:centering}.

\setcounter{table}{4}
\begin{table}
\caption{Lattice centerings and the reflection conditions applied to
conventional Miller indices, where $n$ is an integer.}
\label{tab:centering}
\begin{tabular}{lcl}
Centering & Symbol & Condition \\
\hline\hline
Primitive & $P$ & none \\
$A$-face centered & $A$ & $k+l=2n$ \\
$B$-face centered & $B$ & $h+l=2n$ \\
$C$-face centered & $C$ & $h+k=2n$ \\
Body centered & $I$ & $h+k+l=2n$ \\
All-face centered & $F$ & $h+k,\ h+l,\ k+l=2n$ \\
Rhombohedral (obverse) & $R$ & $-h+k+l=3n$ \\
\hline
\end{tabular}
\end{table}

\section{Comparison of optimizers}
\label{app:optimizer}

The global search could use any population-based optimizer that needs neither
gradients nor an initial guess. To choose one, differential evolution was
compared with particle swarm optimization \cite{kennedy_particle_1995} under
identical conditions, with the same informative peaks and coarse subset, objective,
$k$-d tree, worker pool and subsequent assignment and Wahba refinement
(Algorithm~1) and only the optimizer replaced. The test problems were six
synthetic T4 lysozyme patterns at random orientations ($2\le\lambda\le10$~\AA,
300 peaks, $0.15^\circ$ angular noise) and the measured pattern of Table~\ref{tab:accuracy}, with the refined solution
as reference. Each combination of problem, optimizer and budget was searched
once, without restarts, with five seeds. A search succeeded if the refined
orientation was within $1^\circ$ of the reference, up to the symmetry of the
lattice, and $\log_{10}p<-5$. Particle swarm used a global-best swarm with as
many particles as the differential-evolution population, which SciPy's Sobol
initialization rounds up to a power of two (256 for a nominal 180 and 1024 for
900), so that both optimizers made the same number of objective evaluations,
with four combinations of
the inertia $w$ and the acceleration coefficients $c_1=c_2=c$, while differential evolution used the settings of Section~\ref{sec:search}.

Table~\ref{tab:optimizer} lists the successes. At the smallest budget the best
particle-swarm setting succeeded more often than differential evolution (16
against 9 of 35), and at the intermediate budget about as often (20 against
19). At the largest tested budget, which uses the default population size,
differential evolution succeeded in 94\% of searches against 83\% for the best particle swarm.
Particle swarm was also sensitive to its parameters, with 34--83\% success at
the largest budget depending on the setting, whereas differential evolution
used one parameter setting for all test problems.

The difference is consistent with how the two optimizers keep the population
diverse. The objective has a narrow basin at the true orientation and many
local minima where chance matches accumulate. In global-best particle swarm
every particle is drawn toward the single best position found so far, so the
swarm can contract onto one of these false minima before the true basin is
sampled. With the \texttt{rand/1/bin} strategy, differential evolution has no
such common attractor. Each trial vector is built from randomly chosen members
and replaces only its own parent, and its steps, scaled differences between
members, shrink only as the population itself contracts. In line with this,
particle swarm did better with the more exploratory settings, larger inertia
and acceleration coefficients (Table~\ref{tab:optimizer}). Its faster
contraction is an advantage when few evaluations are available, while at the
largest tested budget the results are consistent with more frequent trapping
in false minima.

\setcounter{table}{5}
\begin{table}
\caption{Successful searches out of 35 (seven problems, five seeds) for
differential evolution (DE) and particle swarm optimization (PSO) at three
budgets (nominal population per parameter $\times$ generations), with the
number of objective evaluations, equal for both optimizers, below each budget
and the success rate in parentheses.}
\label{tab:optimizer}
\centering
\begin{tabular}{lccc}
Optimizer & $60\times100$ & $60\times300$ & $300\times300$ \\
 & (26k) & (77k) & (308k) \\
\hline\hline
DE & 9 (26\%) & 19 (54\%) & 33 (94\%) \\
PSO, $w=0.5$, $c=1.0$ & 6 (17\%) & 8 (23\%) & 12 (34\%) \\
PSO, $w=0.5$, $c=1.49$ & 7 (20\%) & 10 (29\%) & 21 (60\%) \\
PSO, $w=0.72$, $c=1.0$ & 12 (34\%) & 13 (37\%) & 20 (57\%) \\
PSO, $w=0.72$, $c=1.49$ & 16 (46\%) & 20 (57\%) & 29 (83\%) \\
\hline
\end{tabular}
\end{table}

\section{Chance-match model}
\label{app:chance}

This appendix gives the details behind the interpretation of the chance-match
score in Section~\ref{sec:significance}.

For a sum of independent Bernoulli variables with probabilities $p_i$, the
error of a Poisson approximation with mean $\Lambda=\sum_ip_i$ is bounded by
$\sum_ip_i^2$ \cite{le_cam_approximation_1960}, so the approximation is best
for peaks with small $p_i$. For large cells at a wide band this bound is
uninformative ($\sum_ip_i^2\approx60$ for the T4 lysozyme peaks). For
independent matches the Poisson variance $\Lambda$ would exceed the Bernoulli
variance $\sum_ip_i(1-p_i)$ and widen the tail, making the score conservative,
but because the matches are correlated this is not guaranteed.

The number of distinguishable orientations follows from the density
$(1-\cos\omega)/\pi$ of rotation angles \cite{miles_random_1965}. Integrating
it up to $\tau$ gives the probability $\tau^3/6\pi$ that a random rotation lies
within $\tau$ of a given one, so orientation space contains about $6\pi/\tau^3$
distinguishable orientations, $\sim10^8$ at $\tau=0.35^\circ$, comparable to
the $10^8$ orientations of the $0.6^\circ$ grid used for library-based Laue
indexing \cite{sharma_lauematching_2026}. Multiplying $p$ by this count
estimates the expected number of chance solutions with that score anywhere in
orientation space, so $p$ must be well below $10^{-8}$ before such a solution
becomes unlikely. It treats the orientations as independent trials,
which an adaptive search followed by refinement is not, so it is a heuristic
rather than a rigorous bound.

For the 351 informative T4 lysozyme peaks, the $10^5$ random orientations give
a mean of 140.5 indexed peaks and a variance of 89, against $\Lambda=132.3$ and
a Poisson variance of 132. A normal distribution with the observed mean and
variance has a larger upper tail than Eq.~(\ref{eq:pvalue}) below about 170
indexed peaks and a smaller one above. The acceptance threshold,
$\log_{10}p=-5$, corresponds to about 184 peaks, above this crossover, which
suggests that the score is conservative near the threshold. Because $10^5$
orientations do not sample the far tail, this is not a guarantee.



\ack{\noindent\textbf{Acknowledgements}}

The authors would like to acknowledge Aaron Celestian, Minerals Curator at the
Natural History Museum of Los Angeles County, who provided the mesolite single
crystal sample used as a standard crystal for diffraction comparison in this
work.

This material is based upon work supported by the US
Department of Energy, Office of Science, Office of Basic
Energy Sciences Biopreparedness Research Virtual Environment (BRaVE) program.

NW and IY acknowledge support through the ORNL Next Generation Pathways to
Computing (NextGen) program.
FH acknowledges the ORNL Next Generation STEM Internship (NGSI) program.

This research used resources at the High Flux Isotope
Reactor and Spallation Neutron Source, a DOE Office of Science User Facility
operated by the Oak Ridge National Laboratory. The beam time was allocated to
IMAGINE
and MANDI on proposal numbers IPTS-8776, 34720 and 37331.

This manuscript has been authored by UT-Battelle, LLC, under contract
DE-AC05-00OR22725 with the US Department of Energy (DOE). The US government
retains and the publisher, by accepting the article for publication,
acknowledges that the US government retains a nonexclusive, paid-up,
irrevocable, worldwide license to publish or reproduce the published form of
this manuscript, or allow others to do so, for US government purposes. DOE will
provide public access to these results of federally sponsored research in
accordance with the DOE Public Access Plan
(\url{https://www.energy.gov/doe-public-access-plan}).

\bibliography{macro}

\end{document}